\documentclass[fleqn,usenatbib]{mnras}

\usepackage{newtxtext,newtxmath}

\usepackage[T1]{fontenc}

\DeclareRobustCommand{\VAN}[3]{#2}
\let\VANthebibliography\thebibliography
\def\thebibliography{\DeclareRobustCommand{\VAN}[3]{##3}\VANthebibliography}

\usepackage{graphicx}	
\usepackage{amsmath}	
\usepackage{caption}
\usepackage{subcaption}
\usepackage{upgreek}
\usepackage{longtable}
\usepackage{arydshln}
\usepackage{booktabs}

\newcommand{\rone}{FRB~20121102A}

\newcommand{\frb}{FRB~20200120E}
\newcommand{\sgr}{SGR~1935$+$2154}

\title[\frb\ burst storm]{A burst storm from the repeating \frb\ in an M81 globular cluster}

\author[K. Nimmo et al.]{K. Nimmo,$^{1,2}$\thanks{E-mail: k.nimmo@uva.nl}
J. W. T. Hessels,$^{2,1}$
M. P. Snelders,$^{1,2}$
R. Karuppusamy,$^{3}$
D. M. Hewitt,$^{2}$
F. Kirsten,$^{1,4}$\newauthor
B. Marcote,$^{5}$
U. Bach,$^{3}$
A. Bansod,$^{3}$
E. D. Barr,$^{3}$
J. Behrend,$^{3}$
V. Bezrukovs,$^{6}$
S. Buttaccio,$^{7}$
R. Feiler,$^{8}$\newauthor
M. P. Gawro\'nski,$^{8}$
M. Lindqvist,$^{4}$
A. Orbidans,$^{6}$
W. Puchalska,$^{8}$
N. Wang,$^{9}$
T. Winchen,$^{3}$
P. Wolak,$^{8}$\newauthor
J. Wu,$^{3}$
and J.~Yuan$^{9}$
\\
$^{1}$ASTRON, Netherlands Institute for Radio Astronomy, Oude Hoogeveensedijk 4, 7991 PD Dwingeloo, The Netherlands\\
$^{2}$Anton Pannekoek Institute for Astronomy, University of Amsterdam, Science Park 904, 1098 XH, Amsterdam, The Netherlands\\
$^{3}$Max-Planck-Institut f\"{u}r Radioastronomie, Auf dem H\"{u}gel 69, D-53121 Bonn, Germany\\
$^{4}$Department of Space, Earth and Environment, Chalmers University of Technology, Onsala Space Observatory, 439 92, Onsala, Sweden\\
$^{5}$Joint Institute for VLBI ERIC, Oude Hoogeveensedijk 4, 7991 PD Dwingeloo, The Netherlands\\
$^{6}$Engineering Research Institute Ventspils International Radio Astronomy Centre (ERI VIRAC) of Ventspils University of Applied Sciences (VUAS),\\ Inzenieru street 101, Ventspils, LV-3601, Latvia\\
$^{7}$Istituto Nazionale di Astrofisica, Istituto di Radioastronomia Radiotelescopio di Noto, Noto, Italy\\
$^{8}$Institute of Astronomy, Faculty of Physics, Astronomy and Informatics, Nicolaus Copernicus University, Grudziadzka 5, 87-100, Toru\'n, Poland\\
$^{9}$Xinjiang Astronomical Observatory, Chinese Academy of Sciences, 150 Science 1-Street, Urumqi, Xinjiang 830011, People’s Republic of China
}

\date{Accepted 2023 January 21. Received 2023 January 20; in original form 2022 June 8}

\pubyear{2022}

\begin{document}
\label{firstpage}
\pagerange{\pageref{firstpage}--\pageref{lastpage}}
\maketitle

\begin{abstract}
The repeating fast radio burst (FRB) source \frb\ is exceptional because of its proximity and association with a globular cluster. Here we report $60$ bursts detected with the Effelsberg telescope at 1.4\,GHz. We observe large variations in the burst rate, and report the first \frb\ `burst storm', where the source suddenly became active and 53 bursts (fluence $\geq 0.04$\,Jy\,ms) occurred within only 40 minutes. We find no strict periodicity in the burst arrival times, nor any evidence for periodicity in the source's activity between observations. The burst storm shows a steep energy distribution (power-law index $\alpha = 2.39\pm0.12$) and a bi-modal wait-time distribution, with log-normal means of 0.94$^{+0.07}_{-0.06}$\,s and 23.61$^{+3.06}_{-2.71}$\,s. We attribute these wait-time distribution peaks to a characteristic event timescale and pseudo-Poisson burst rate, respectively. The secondary wait-time peak at $\sim1$\,s is $\sim50\times$ longer than the $\sim24$\,ms timescale seen for both FRB~20121102A and FRB~20201124A --- potentially indicating a larger emission region, or slower burst propagation. \frb\ shows order-of-magnitude lower burst durations and luminosities compared with FRB~20121102A and FRB~20201124A. Lastly, in contrast to FRB~20121102A, which has observed dispersion measure (DM) variations of $\Delta{\rm DM} >1$\,pc\,cm$^{-3}$ on month-to-year timescales, we determine that \frb's DM has remained stable ($\Delta{\rm DM} <0.15$\,pc\,cm$^{-3}$) over $>10$\,months. Overall, the observational characteristics of \frb\ deviate quantitatively from other active repeaters, but it is unclear whether it is qualitatively a different type of source.
\end{abstract}

\begin{keywords}
Fast Radio Bursts 
\end{keywords}



\section{Introduction}

\label{sec:intro}
Fast radio bursts (FRBs) are highly luminous, millisecond-duration radio transients, originating at extragalactic distances \citep{lorimer_2007_sci,thornton_2013_sci}. Despite 15 years of research in the field (for recent reviews see, e.g., \citealt{petroff_2019_aarv,petroff_2022_aarv,cordes_2019_araa}), and more than $600$ FRBs discovered to date (e.g., \citealt{chime_2021_apjs}), their nature is still debated. The diverse burst phenomenology \citep{pleunis_2021_apj}, including a relatively small fraction ($\sim4$\%) of FRB sources exhibiting repeating bursts \citep{spitler_2016_natur}, potentially indicates multiple FRB origins. Athough FRBs are highly luminous, their large extragalactic distances (typically hundreds of megaparsecs to gigaparsecs) mean that we are strongly sensitivity limited, and therefore only observe the bright end of the distribution of potentially observable fast radio transients.

FRBs in the local Universe (luminosity distance $d_L <$ a few hundred Mpc) provide us with the unique opportunity to connect our knowledge of fast radio transients in the Milky Way and its satellites --- e.g., the Crab pulsar \citep{hankins_2007_apj}, the `Crab twin' in the Large Magellanic Cloud PSR~B0540$-$69 \citep{geyer_2021_mnras} and the Galactic magnetar SGR~1935+2154 \citep{chime_2020_natur_587,bochenek_2020_natur} --- to the much more distant FRB population. We can do this through detailed characterisation of their local environments (e.g., \citealt{marcote_2020_natur, tendulkar_2021_apjl, kirsten_2022_natur}), by applying strong constraints on multi-wavelength counterparts to the radio emission \citep{scholz_2020_apj}, and by conducting higher-sensitivity searches for low-luminosity FRBs \citep{kirsten_2022_natur,nimmo_2022_natas,majid_2021_apjl}.

The closest known extragalactic FRB discovered to date, \frb\ \citep{bhardwaj_2021_apjl}, was discovered by the Canadian Hydrogen Intensity Mapping Experiment FRB project (CHIME/FRB; \citealt{chime_2018_apj}) and subsequently precisely localised using the European Very long baseline interferometry (VLBI) Network (EVN) to a globular cluster in the M81 galactic system \citep{kirsten_2022_natur}. Not only is the globular cluster origin of \frb\ in stark contrast to the star-forming environments of other well-studied repeating FRBs \citep{chatterjee_2017_natur,marcote_2017_apjl,bassa_2017_apjl,marcote_2020_natur,tendulkar_2021_apjl,nimmo_2022_apjl,fong_2021_apjl,ravi_2022_mnras}, but the luminosities of the bursts are 1--2 orders of magnitude weaker than those observed from other repeaters, and even less luminous than the bright FRB-like transient from \sgr\ \citep{chime_2020_natur_587,bochenek_2020_natur}. Furthermore, the \frb\ burst widths are atypically narrow (a factor of $\sim 30$ shorter than typical FRB~20121102A bursts at a comparable frequency; \citealt{nimmo_2022_natas, majid_2021_apjl, li_2021_natur}). 

\citet{nimmo_2022_natas} discuss the observational connections between \frb, the Crab pulsar, the Galactic magnetar SGR~1935+2154, and the more distant FRBs using the measured luminosities, range of burst timescales, burst morphologies, and polarimetry. In doing so, they highlight the spectrum of short-duration radio emission spanning many orders of magnitude in luminosity and timescales, and emphasise that the exact division between transient types (e.g., pulsar and FRB emission) is presently unclear.

To date, only $14$ bursts from \frb\ have been reported in the literature \citep{bhardwaj_2021_apjl,nimmo_2022_natas,majid_2021_apjl}\footnote{Additional CHIME/FRB bursts are reported on in the \href{https://www.chime-frb.ca/repeaters/FRB20200120E}{CHIME/FRB public database.}}. A larger sample of bursts from \frb\ provides the ability to probe its time-variable activity, search for any underlying periodicity and study the energy and wait-time distributions to compare with similar studies of Galactic neutron stars, and other repeating FRBs. 

Here we report the detection of $60$ new bursts from \frb, detected via monitoring with the 100-m Effelsberg telescope at 1.4\,GHz from 2021 April to 2022 April. During this monitoring, we report the first observed `burst storm' from \frb\, where $53$ of the $60$ bursts occurred within a $\sim40$\,minute time window. In Section~\ref{sec:observations} we describe the Effelsberg monitoring observations, and the data products. In Section~\ref{sec:search} we describe the search for bursts, followed by a description of the burst analysis and results in Section~\ref{sec:analysis}. In Section~\ref{sec:discussion}, we discuss our results in the context of previous FRB observations, and compare with observations of Galactic neutron stars, before presenting the conclusions of this work in Section~\ref{sec:conclusion}.

\section{Observations}
\label{sec:observations}

\begin{figure*}
\centering
        {\includegraphics[height=80mm,trim=1cm 0cm 1cm 1cm, clip=true]{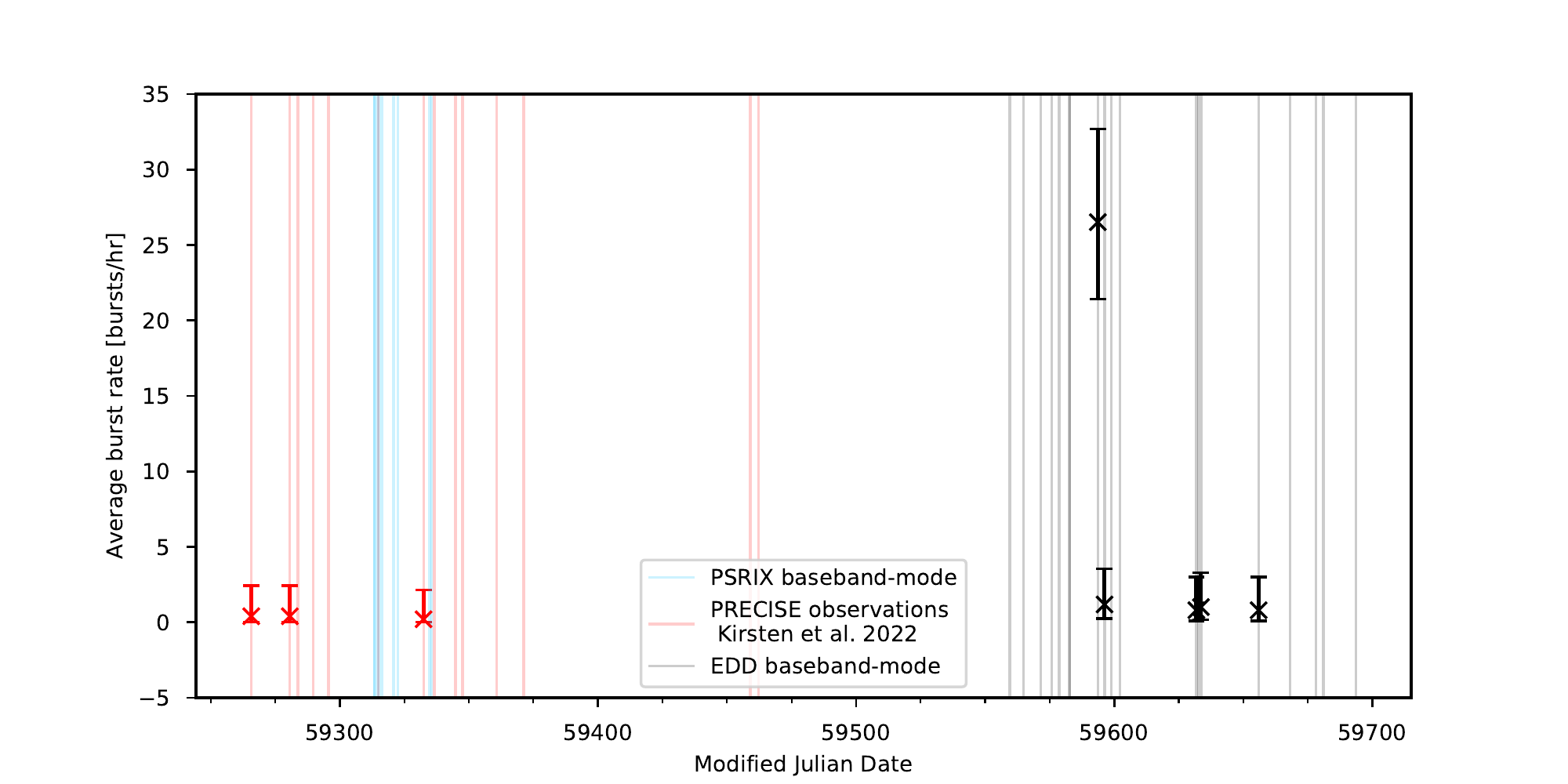}}
  \caption{Summary of the observations of \frb\ presented in this work, in addition to those reported in \citet{kirsten_2022_natur} and \citet{nimmo_2022_natas}. The coloured lines show the observation epochs (where colour signifies the observation type, as described in Section\,\ref{sec:observations}). The cross markers show the average burst rate per observing epoch, unless there were no detections. For a tabular version of this information, see Table\,\ref{tab:observations}.}

     \label{fig:obs}
\end{figure*}

Between December 11 2021 and April 24 2022, we monitored \frb\ (using the EVN-PRECISE\footnote{\href{https://www.ira.inaf.it/precise/Home.html}{Pinpointing Repeating
Chime Sources with the EVN}} position; \citealt{kirsten_2022_natur}) with the 100-m Effelsberg telescope using the recently developed Effelsberg Direct Digitization (EDD) backend operating in baseband-mode. This allowed us to simultaneously record total intensity pulsar backend {\tt psrfits} \citep{hotan_2004_pasa} data with time and frequency resolution of 40.96\,$\upmu$s and 0.1953\,MHz, respectively (exceptions to this are noted in Table\,\ref{tab:observations}), alongside the raw voltages (`baseband' data, dual circular polarisation) in Data Acquisition and Distributed Analysis (DADA) format \citep{vanstraten_2021_ascl} sampled at 1/$400$\,MHz. We used the central pixel of the 7-beam 21-cm receiver, with an observing bandwidth from 1.2--1.6\,GHz. In total, we have observed for 28.4\,hr using this observing setup, which is summarised in Table\,\ref{tab:observations}. We recorded {\tt psrfits} data of the test pulsar B0355+54. Due to an incorrect observing set-up, the raw voltages of 2022 February 21 and 22 (MJDs 59631, 59632) were not recorded. 

We add the EVN-PRECISE observations originally reported in \citet{kirsten_2022_natur} and \citet{nimmo_2022_natas} to Table\,\ref{tab:observations}, with an additional three PRECISE observations using the VLBI backend that occurred on 2021 June 6, September 2 and September 5. Using the VLBI Digital Base Band Converter (DBBC2; \citealt{tuccari_2010_ivs}) backend at Effelsberg, these observations recorded baseband data with dual circular polarisation (2-bit sampling) in VDIF \citep{whitney_2010_ivs} format. Additionally, we recorded total intensity filterbank data with the PSRIX pulsar backend \citep{lazarus_2016_mnras} with a time and frequency resolution of 102.4\,$\upmu$s and 0.49\,MHz, respectively. The observing bandwidth is 1255--1505\,MHz and the total observing time using this observing setup is 63.6\,hr (all but 12.1\,hr originally reported in \citealt{kirsten_2022_natur} and \citealt{nimmo_2022_natas}). Note that these are VLBI observations, and therefore had frequent scans of calibrator sources. In such cases, the time on source is therefore approximately 65\,\% of the reported times.

Furthermore, we also report on observations obtained through Director's Discretionary Time at Effelsberg between 2021 April 9 and 2021 May 1. These observations were using the PSRIX pulsar backend in baseband-mode: recording total intensity filterbank data products with time and frequency resolution of 65.5\,$\upmu$s and 0.24\,MHz, respectively, simultaneously with the raw voltages (dual circular). The observing bandwidth is 1233--1483\,MHz, and total observing time with this setup is 13.7\,hr.

All of the observations are summarised in Table\,\ref{tab:observations} and Figure\,\ref{fig:obs}. 

\begin{table*}
\caption{\label{tab:observations} \frb\ Effelsberg monitoring observation details.}
\centering{\begin{tabular}{ccccc}
\hline
\hline
{Start MJD$\mathrm{^{a}}$} & {Duration [hr]} & {Observation type} & {Number of bursts} & {Average burst rate [/hr]}\\ 
\hline  
{59265.708$\mathrm{^{b}}$} & {4.99} & {PRECISE, VLBI backend} & {2} & {$0.4^{+2.0}_{-0.4}$} \\ 
 {59280.656$\mathrm{^{b}}$} & {4.99} & {PRECISE, VLBI backend} & {2} & {$0.4^{+2.0}_{-0.4}$} \\ 
 {59283.792$\mathrm{^{b}}$} & {4.99} & {PRECISE, VLBI backend} & {0} & {$0.0^{+1.8}_{-0.0}$} \\ 
 {59289.750$\mathrm{^{b}}$} & {4.99} & {PRECISE, VLBI backend} & {0} & {$0.0^{+1.8}_{-0.0}$} \\ 
 {59295.667$\mathrm{^{b}}$} & {4.99} & {PRECISE, VLBI backend} & {0} & {$0.0^{+1.8}_{-0.0}$} \\ 
 {59313.437} & {0.07} & {PSRIX baseband-mode} & {0} & {$0.0^{+1.8}_{-0.0}$} \\ 
 {59313.458} & {1.00} & {PSRIX baseband-mode} & {0} & {$0.0^{+1.8}_{-0.0}$} \\ 
 {59314.508} & {2.21} & {PSRIX baseband-mode} & {0} & {$0.0^{+1.8}_{-0.0}$} \\ 
 {59314.887$\mathrm{^{b}}$} & {2.04} & {PRECISE, VLBI backend} & {0} & {$0.0^{+1.8}_{-0.0}$} \\ 
 {59315.191} & {0.89} & {PSRIX baseband-mode} & {0} & {$0.0^{+1.8}_{-0.0}$} \\ 
 {59316.235} & {1.17} & {PSRIX baseband-mode} & {0} & {$0.0^{+1.8}_{-0.0}$} \\ 
 {59320.828} & {2.22} & {PSRIX baseband-mode} & {0} & {$0.0^{+1.8}_{-0.0}$} \\ 
 {59322.514} & {2.00} & {PSRIX baseband-mode} & {0} & {$0.0^{+1.8}_{-0.0}$} \\ 
 {59332.458$\mathrm{^{b}}$} & {4.99} & {PRECISE, VLBI backend} & {1} & {$0.2^{+1.9}_{-0.2}$} \\ 
 {59334.807} & {2.08} & {PSRIX baseband-mode} & {0} & {$0.0^{+1.8}_{-0.0}$} \\ 
 {59335.634} & {2.01} & {PSRIX baseband-mode} & {0} & {$0.0^{+1.8}_{-0.0}$} \\ 
 {59336.708$\mathrm{^{b}}$} & {7.01} & {PRECISE, VLBI backend} & {0} & {$0.0^{+1.8}_{-0.0}$} \\ 
 {59344.771$\mathrm{^{b}}$} & {2.50} & {PRECISE, VLBI backend} & {0} & {$0.0^{+1.8}_{-0.0}$} \\ 
 {59347.625$\mathrm{^{b}}$} & {4.99} & {PRECISE, VLBI backend} & {0} & {$0.0^{+1.8}_{-0.0}$} \\ 
 {59360.708$\mathrm{^{b}}$} & {4.99} & {PRECISE, VLBI backend} & {0} & {$0.0^{+1.8}_{-0.0}$} \\ 
 {59371.234} & {2.48} & {PRECISE, VLBI backend} & {0} & {$0.0^{+1.8}_{-0.0}$} \\ 
 {59459.027} & {4.50} & {PRECISE, VLBI backend} & {0} & {$0.0^{+1.8}_{-0.0}$} \\ 
 {59462.168} & {5.11} & {PRECISE, VLBI backend} & {0} & {$0.0^{+1.8}_{-0.0}$} \\ 
 {59559.584$\mathrm{^{c}}$} & {2.00} & {EDD baseband-mode} & {0} & {$0.0^{+1.8}_{-0.0}$} \\ 
 {59564.902} & {2.00} & {EDD baseband-mode} & {0} & {$0.0^{+1.8}_{-0.0}$} \\ 
 {59571.573} & {2.00} & {EDD baseband-mode} & {0} & {$0.0^{+1.8}_{-0.0}$} \\ 
 {59575.735} & {1.38} & {EDD baseband-mode} & {0} & {$0.0^{+1.8}_{-0.0}$} \\ 
 {59578.622} & {2.00} & {EDD baseband-mode} & {0} & {$0.0^{+1.8}_{-0.0}$} \\ 
 {59582.677} & {0.20} & {EDD baseband-mode} & {0} & {$0.0^{+1.8}_{-0.0}$} \\ 
 {59582.785} & {0.50} & {EDD baseband-mode} & {0} & {$0.0^{+1.8}_{-0.0}$} \\ 
 {59593.650} & {2.00} & {EDD baseband-mode} & {53} & {$26.5^{+6.2}_{-5.1}$} \\
 {59596.262} & {0.85} & {EDD baseband-mode} & {1} & {$1.2^{+2.4}_{-0.9}$} \\ 
 {59598.871} & {2.50} & {EDD baseband-mode} & {0} & {$0.0^{+1.8}_{-0.0}$} \\ 
 {59602.235} & {0.03} & {EDD baseband-mode} & {0} & {$0.0^{+1.8}_{-0.0}$} \\ 
 {59631.818$\mathrm{^{d}}$} & {2.50} & {EDD baseband-mode} & {2} & {$0.8^{+2.2}_{-0.7}$} \\ 
 {59632.673$\mathrm{^{d}}$} & {2.15} & {EDD baseband-mode} & {0} & {$0.0^{+1.8}_{-0.0}$} \\ 
 {59633.599} & {1.99} & {EDD baseband-mode} & {2} & {$1.0^{+2.3}_{-0.8}$} \\ 
 {59655.950} & {2.50} & {EDD baseband-mode} & {2} & {$0.8^{+2.2}_{-0.7}$} \\ 
 {59668.052} & {0.78} & {EDD baseband-mode} & {0} & {$0.0^{+1.8}_{-0.0}$} \\ 
 {59678.079} & {1.36} & {EDD baseband-mode} & {0} & {$0.0^{+1.8}_{-0.0}$} \\ 
 {59680.974} & {1.65} & {EDD baseband-mode} & {0} & {$0.0^{+1.8}_{-0.0}$} \\ 
 {59693.655} & {0.85} & {EDD baseband-mode} & {0} & {$0.0^{+1.8}_{-0.0}$} \\ 
 \hline   

\multicolumn{4}{l}{$\mathrm{^{a}}$ Topocentric.}\\
\multicolumn{4}{l}{$\mathrm{^{b}}$ Originally reported in \cite{kirsten_2022_natur} and \cite{nimmo_2022_natas}.}\\
\multicolumn{4}{l}{$\mathrm{^{c}}$ Frequency resolution of these data is 0.4\,MHz.}\\
\multicolumn{4}{l}{$\mathrm{^{d}}$ No raw voltages due to incorrect observing set-up.}\\
\end{tabular}} 
\end{table*} 

\section{Burst search and discovery}
\label{sec:search}
\subsection{EDD baseband-mode}
The total intensity EDD {\tt psrfits} data were first converted to filterbank format using {\tt digifil} \citep{vanstraten_2011_pasa}, at the native resolution of the data, to be compatible with {\tt Heimdall}\footnote{\tt https://sourceforge.net/projects/heimdall-astro/}. Frequency channels in our observing band that frequently contain radio frequency interference (RFI) were masked before searching for single pulses with {\tt Heimdall}, using a S/N threshold of 7. Candidates found in the {\tt Heimdall} search were then classified using FETCH (models A and H, probability threshold 0.5; \citealt{agarwal_2020_mnras}). We inspected the FETCH candidate plots by eye, and also the plots of any {\tt Heimdall} candidate within $\pm3$\,pc\,cm$^{-3}$ of the known FRB dispersion measure (DM) and above a S/N threshold of 10. 

In this search, we found 37 bursts on 2022 January 14 (MJD 59593), 1 burst on 2022 January 17 (MJD 59596), and 2 bursts on each of 2022 February 21, 23 and March 17 (MJDs 59631, 59633, and 59655, respectively). On closer inspection of the {\tt Heimdall} candidates on 2022 January 14 (pre-FETCH), with DMs $\pm1$\,pc\,cm$^{-3}$ around the best-known value of 87.7527\,pc\,cm$^{-3}$ \citep{nimmo_2022_natas}, an additional 7 bursts were discovered. These bursts were all reported with relatively low {\tt Heimdall} S/N values of $<10$. Therefore the low S/N, combined with the narrow temporal burst widths, are likely the cause of the misclassification by FETCH. A similar exercise was repeated on the other observations, and no additional bursts were discovered. Note that post 2022 January 14, we altered the analysis pipeline to keep candidate plots for inspection for any Heimdall candidate within $\pm3$\,pc\,cm$^{-3}$ of the known FRB DM and above a S/N threshold of  7. 

Due to the high density of bursts discovered on 2022 January 14, we saved the raw voltages for the entire 2\,hr observation for further inspection. This was not possible for all observations due to the high data volume of the raw voltages (1\,hr is approximately 5.5\,TB of raw voltage data). Therefore, for the bursts discovered at other epochs, we saved only the 2.048\,s baseband recording containing the burst (and sometimes also the neighbouring recording if the dispersion sweep crossed between recordings).

\begin{table*}
\caption{\label{tab:search}The success of burst searches on 2022 January 14 EDD data. \checkmark\ indicates that a burst was classified as astrophysical by FETCH (models A and H, probability threshold 0.5). The S/N values quoted for both {\tt Heimdall} and {\tt PRESTO} searches are boxcar S/N ratios.
}
\centering\hspace{-5em}{\begin{tabular}{lrcrrr}
\hline
\hline
\multicolumn{1}{c}{} & \multicolumn{2}{c}{$40.96$\,$\upmu$s pulsar} & \multicolumn{2}{c}{$40.96$\,$\upmu$s baseband} & \multicolumn{1}{c}{$1.28$\,$\upmu$s baseband} \\
{Burst} & {{\tt Heimdall} S/N} & {FETCH} & {{\tt Heimdall} S/N} &  {{\tt PRESTO} S/N} & {{\tt PRESTO} S/N} \\
\hline
{B1} & {13.6} & {\checkmark} & {12.4} & {13.4} & {9.3} \\\cdashline{4-6}
{B2} & {36.6} & {\checkmark} & \multicolumn{3}{c}{Lost baseband data} \\\cdashline{4-6}
{B3} & {12.8} & {\checkmark} & {10.8} & {10.8} & {--}\\
{B4} & {7.9} & {--} & {7.5} & {8.6}  & {--}\\
{B5} & {21.6} & {\checkmark} & {17.8} & {21.6} & {9.6} \\
{B6} & {--} & {--} & {7.3} & {--} & {--} \\
{B7} & {8.7} & {\checkmark} & {8.7} & {9.6} & {--} \\
{B8} & {9.3} & {\checkmark} & {8.4} & {10.3} & {--} \\
{B9} & {8.4} & {\checkmark} & {7.6} & {8.7} & {--} \\
{B10} & {20.2} & {\checkmark} & {19.0} & {20.6} & {12.9}\\
{B11} & {7.8} & {\checkmark} & {--} & {8.7} & {--} \\
{B12} & {16.7} & {\checkmark} & {16.5} & {17.9} &  {14.7}\\
{B13} & {20.8} & {\checkmark} & {16.7} & {18.6} &  {14.5}\\
{B14} & {--} & {--} & {9.7} & {10.9} & {8.7} \\
{B15} & {--} & {--} & {--} & {--} & {7.7}\\
{B16} & {7.2} & {--} & {7.7} & {8.0} & {7.0} \\
{B17} & {24.7} & {\checkmark} & {22.3} & {25.0} &  {17.2}\\
{B18} & {--} & {--} & {11.9} & {13.1} &  {10.6}\\
{B19} & {--} & {--} & {7.0} & {7.8} &  {9.6}\\
{B20} & {19.8} & {\checkmark} & {18.8} & {21.7}  & {16.3}\\
{B21} & {17.2} & {\checkmark} & {14.2} & {17.2} & {15.4}\\
{B22} & {7.6} & {\checkmark} & {--} & {7.2} &  {--}\\
{B23} & {11.0} & {\checkmark} & {9.0} & {10.5}  & {9.8}\\
{B24} & {10.2} & {\checkmark} & {9.1} & {10.3}  & {7.4}\\
{B25} & {7.6} & {\checkmark} & {--} & {--}  & {--}\\
{B26} & {17.0} & {\checkmark} & {15.1} & {16.2} & {14.6}\\
{B27} & {7.1} & {\checkmark} & {--} & {7.4} & {--}\\
{B28} & {8.9} & {--} & {--} & {--}  & {7.1}\\
{B29} & {8.1} & {\checkmark} & {7.7} & {8.5} & {--}\\
{B30} & {--} & {--} & {--} & {8.0}  & {--}\\
{B31} & {22.5} & {\checkmark} & {21.1} & {24.3} & {18.3}\\
{B32} & {--} & {--} & {--} & {7.8} &  {9.0}\\
{B33} & {25.2} & {\checkmark} & {19.4} & {26.4}  & {15.3}\\
{B34} & {16.2} & {\checkmark} & {14.4} & {17.3}  & {17.3}\\
{B35} & {21.8} & {\checkmark} & {19.8} & {22.2} & {15.2}\\
{B36} & {8.9} & {--} & {7.4} & {7.9} &  {7.2}\\
{B37} & {20.7} & {\checkmark} & {19.8} & {23.0}  & {25.4}\\
{B38} & {9.5} & {\checkmark} & {7.1} & {11.0}  & {9.5}\\
{B39} & {7.7} & {\checkmark} & {--} & {--}  & {--}\\
{B40} & {12.9} & {\checkmark} & {10.4} & {11.4}  & {10.0}\\
{B41} & {9.2} & {--} & {7.4} & {9.0} & {8.0}\\
{B42} & {8.6} & {--} & {7.0} & {8.0} & {9.2}\\
{B43} & {21.6} & {\checkmark} & {19.5} & {23.3}  & {15.0}\\
{B44} & {15.1} & {\checkmark} & {12.4} & {14.2}  & {12.8}\\
{B45} & {7.1} & {\checkmark} & {--} & {--} & {--}\\
{B46} & {--} & {--} & {--} & {7.2}  & {7.1}\\
{B47} & {20.6} & {\checkmark} & {16.8} & {21.0}  & {18.7}\\
{B48} & {10.8} & {\checkmark} & {9.0} & {11.5}  & {9.4}\\
{B49} & {14.5} & {\checkmark} & {12.1} & {17.1}  & {16.9}\\
{B50} & {14.0} & {\checkmark} & {14.5} & {17.6} & {18.1}\\
{B51} & {8.2} & {\checkmark} & {--} & {9.0}  & {8.7}\\
{B52} & {7.7} & {--} & {--} & {9.2}  & {--}\\
{B53} & {--} & {--} & {--} & {9.7}  & {--}\\

\hline  
\end{tabular}}
\end{table*}

\subsubsection{January 14 baseband data re-search}
\label{sec:research}
Using {\tt digifil}, we created 8-bit total intensity filterbank data from the baseband DADA data at a resolution of 40.96\,$\upmu$s and 0.1953\,MHz in time and frequency, respectively. This resolution matches that of the pulsar data. The goal of searching these data products was to ensure that we understood the filterbank data created and could recover the same bursts discovered in the search of the pulsar data. These data products were searched for single pulses using both {\tt Heimdall} and {\tt PRESTO} \citep{ransom_2001_phdt}. For the {\tt Heimdall}-based search, we mask frequency channels that frequently exhibit RFI before searching for single pulses, and inspect the candidates with a DM within $\pm1$\,pc\,cm$^{-3}$ of the known DM of \frb. For {\tt PRESTO}-based searches, we use the {\tt PRESTO} tool {\tt rfifind} to mask time and frequency blocks that contain RFI before searching for single pulses using {\tt single\_pulse\_search.py}. The {\tt PRESTO} single-pulse candidates were then grouped into events using a modified version of SpS \citep{michilli_2018_mnras}, and events with a DM within $\pm1$\,pc\,cm$^{-3}$ of the known DM were inpected by eye.

The Heimdall search of the baseband data returned $34$ of the $44$ bursts found in the pulsar data: $9$ of those missing all have S/N $<9$, and $1$ (B2) falls within a $\sim4$\,minute gap in the baseband data where we have missing data. The loss of some low-S/N bursts in the baseband search is likely because of different scalings applied in the creation of the filterbank data. We did, however, find $4$ previously undiscovered bursts in the {\tt Heimdall} search of the baseband data. These bursts also have relatively low S/N, and likely were missed in the search of the pulsar data for the same reasons as above. The different scalings applied to create the pulsar data and baseband data products, combined with time-dependent RFI, is likely also the reason for the differing S/N values for the {\tt Heimdall} searches of these data (Table\,\ref{tab:search}).

The {\tt PRESTO} search of the baseband data returned $39$ of the original $44$ bursts, where, again, B2 falls within the data loss region. The other missing $4$ bursts are a subset of the missing bursts in the {\tt Heimdall} search. Of the 4 {\it new} {\tt Heimdall}-discovered bursts, {\tt PRESTO} found $3$. Furthermore, {\tt PRESTO} discovered an additional $4$ low-S/N bursts, bringing the number of bursts discovered in the 2022 January 14 observation to $52$. In general, the {\tt PRESTO} S/N values appear higher than {\tt Heimdall}, which is likely how {\tt PRESTO} found additional bursts that {\tt Heimdall} missed above our detection threshold of $7$. Additionally, the RFI flagging method and boxcar width trialling is different between the {\tt Heimdall} and {\tt PRESTO} searches, which could impact the discovery of bursts, especially in the low-S/N regime. Note there is a timestamp mislabelling in the pulsar backend data, creating a fictitious 125\,ms delay between the baseband recording and pulsar backend recording. This was identified, measured and calibrated for using bursts detected in both the pulsar and baseband data. 

\begin{figure*}
\begin{subfigure}[t]{0.68\linewidth}
    \centering
    \includegraphics[width=\hsize,trim=0cm 0.15cm 0cm 0cm, clip=true]{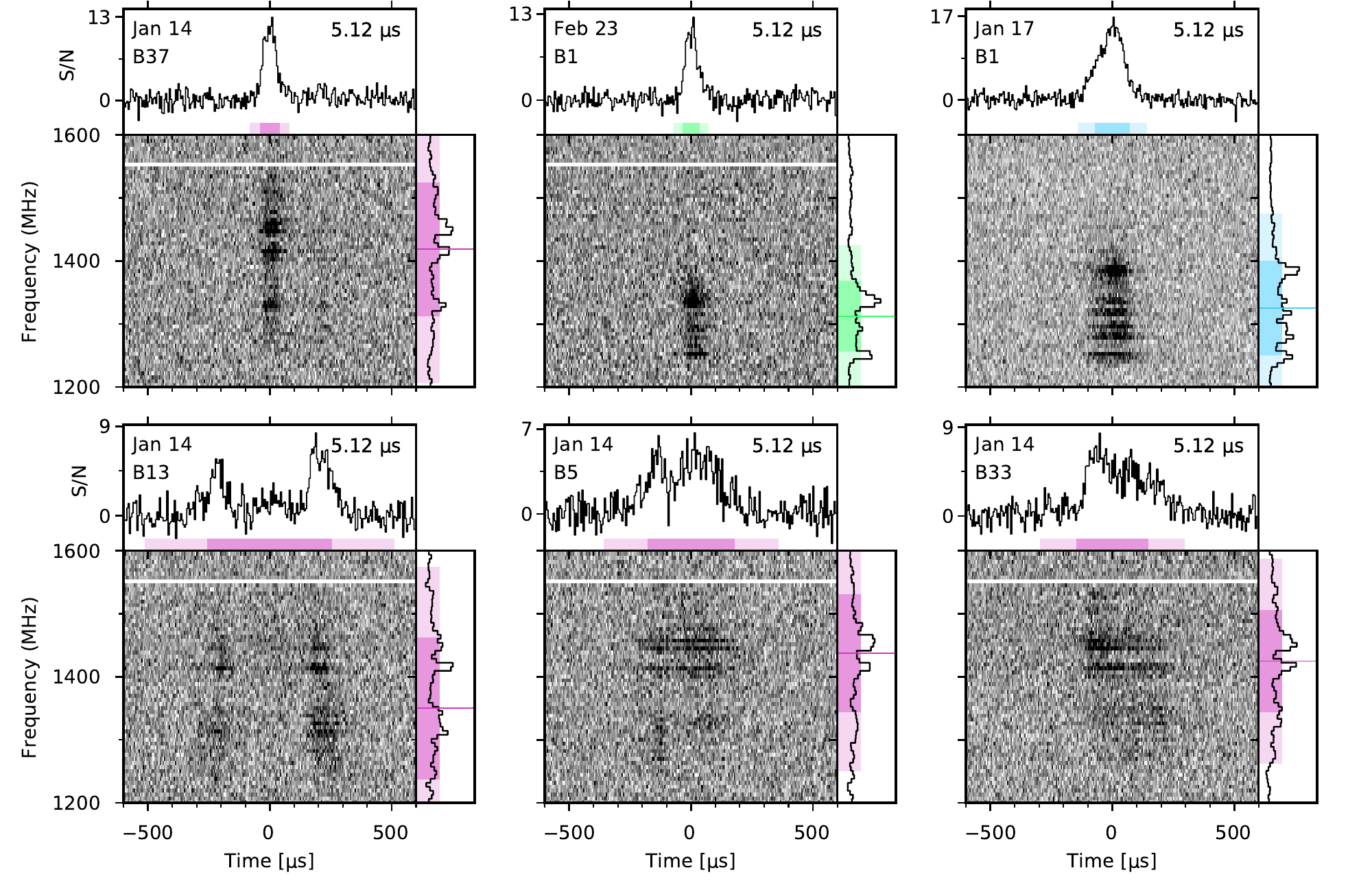}
    \caption{}
    \label{fig:some}
  \end{subfigure}
  \begin{subfigure}[t]{0.31\linewidth}
    \centering
    \includegraphics[width=\hsize,trim=0.15cm 0cm 0cm 0cm, clip=true]{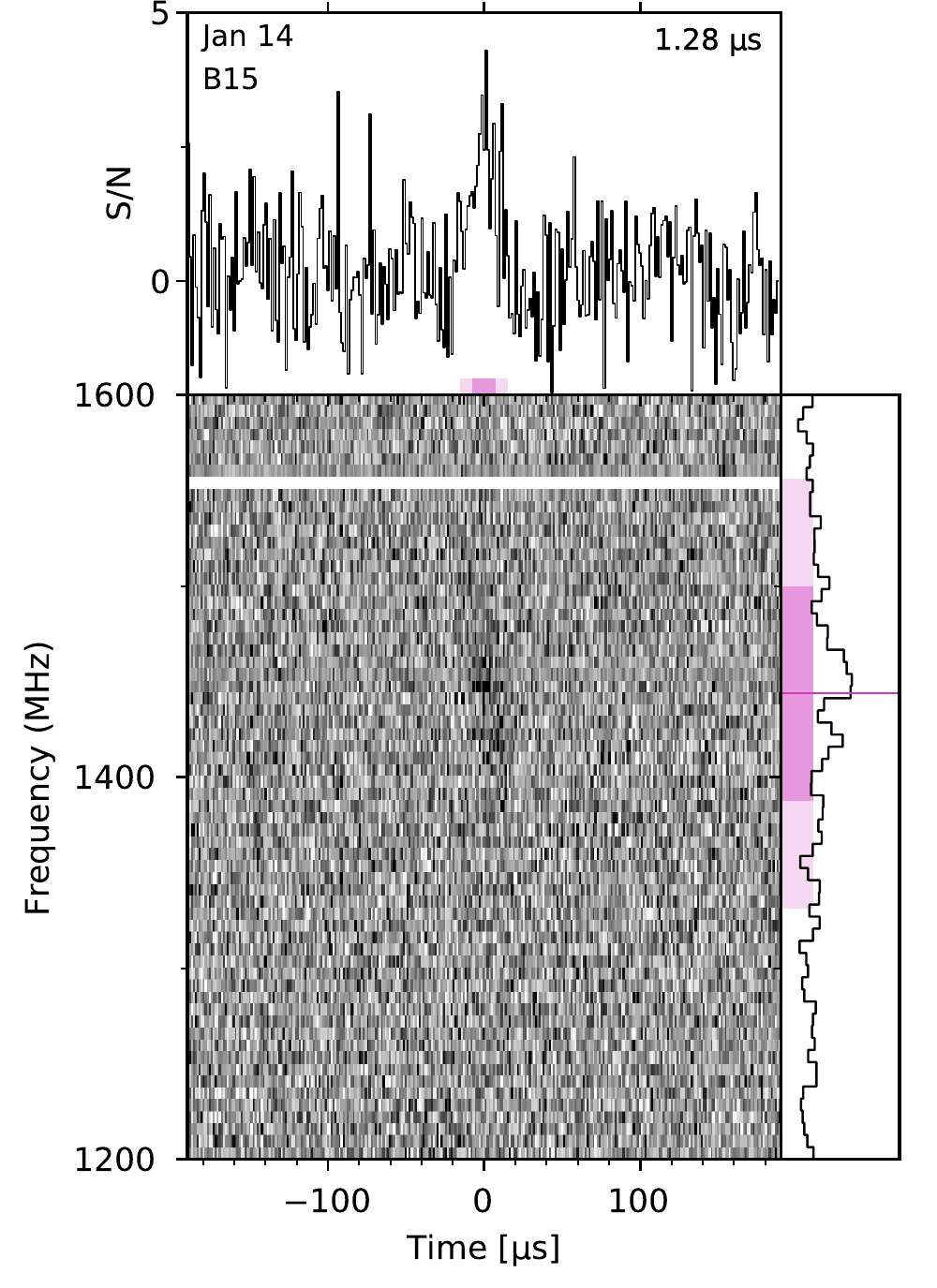}
    \caption{}
    \label{fig:narrow}
  \end{subfigure}
\caption{Burst profiles, dynamic spectra and time-averaged spectra for a subset of the \frb\ bursts presented in this work. The filterbank data plotted is created by channelising the raw voltages to a resolution of 5.12\,$\upmu$s and 0.1953\,MHz in time and frequency, respectively, for sub-figure (a), and 1.28\,$\upmu$s and 0.7813\,MHz for sub-figure (b). Sub-figure (b) shows the burst discovered in the microsecond search of the data (see Section\,\ref{sec:1us}); this is the narrowest burst in this sample. The frequency resolution has been downsampled by a factor of 32 (sub-figure (a)) and 8 (sub-figure (b)) for visualisation purposes. The data are coherently (within frequency channels) and incoherently (between channels) corrected for dispersion using a DM of 87.7527\,pc\,cm$^{-3}$ \citep{nimmo_2022_natas}. The coloured bars highlight the 1-$\sigma$ (dark) and 2-$\sigma$ (light) error regions on the burst extent in both frequency and time, where the different colours represent different observing epochs: 2022 January 14 (purple), 2022 January 17 (blue) and 2022 February 23 (green). Time\,$=0$ represents the burst centroid in time, and the horizontal coloured line in the burst spectrum represents the frequency centroid. Frequency channels that have been masked due to RFI have been omitted from the dynamic spectra (white horizontal lines). The burst profile is created by averaging the dynamic spectrum within the 2-$\sigma$ burst extent in frequency, and similarly the burst spectrum is created by averaging the dynamic spectrum within the 2-$\sigma$ burst extent in time.}  
\label{fig:somebursts}
\end{figure*}

\subsubsection{January 14 1.28\,$\upmu$s search}
\label{sec:1us}
Motivated by the clear microsecond timescales seen in a burst from \frb\ \citep{nimmo_2022_natas}, we re-searched the 2022 January 14 observation at 1.28\,$\upmu$s time resolution. We used {\tt digifil} to create coherently dedispersed (using a DM of $87.7527$\,pc\,cm$^{-3}$; \citealt{nimmo_2022_natas}) 8-bit total intensity filterbank data with time and frequency resolution 1.28\,$\upmu$s and 0.78\,MHz, respectively. The true DM value must be within $\sim0.3$\,pc\,cm$^{-3}$ of the DM used for coherent dedispersion to ensure that the dispersion smearing within a channel is smaller than the time resolution. Fortunately, $0.3$\,pc\,cm$^{-3}$ is much greater than the uncertainty on the known DM \citep{nimmo_2022_natas}, and we confirm in Section\,\ref{sec:analysis} that the DM has not significantly changed from the previous measurement. 

The {\tt Heimdall} search of the 1.28\,$\upmu$s data returned none of the $52$ bursts, and no additional bursts. For the {\tt PRESTO}-based search, we incoherently dedispersed using trial DMs from 87.6 to 87.9\,pc\,cm$^{-3}$ in steps of 0.01\,pc\,cm$^{-3}$. This results in maximal DM step-size smearing of $\sim6$\,$\upmu$s across the entire 400\,MHz band, or $\sim3$\,$\upmu$s for the centre 200\,MHz, where the bursts in our sample reside. We, therefore, downsample to 2.56\,$\upmu$s for the search. The {\tt PRESTO} search returned $35$ of the $52$, and discovered an additional burst, bringing the {\bf total burst count} to 53 on 2022 January 14. The additional burst found in our high-time-resolution search is the narrowest burst in our sample, with a temporal scale of $\sim14$\,$\upmu$s, which combined with its low S/N is the reason it was not caught in either of the 40.96\,$\upmu$s searches. The behaviour of {\tt Heimdall} is not well-studied at extremely high time resolutions (e.g. $\upmu$s) and the fact that we recover a significant fraction of the bursts using PRESTO, implies that {\tt Heimdall} has significantly lost sensitivity at this resolution.

Table\,\ref{tab:search} summarises the results of the searches on the 2022 January 14 observation. The burst profiles, dynamic spectra and time-averaged spectra for selected \frb\ bursts can be found in Figure\,\ref{fig:somebursts}, highlighting the diverse burst morphology observed including the exceptionally narrow $\sim14$\,$\upmu$s burst. In Figure\,\ref{fig:fullburst}(a) and (b) we show the complete time-ordered burst plot for the entire $60$ burst sample presented in this work.

\subsection{PRECISE observations}
\label{sec:PRECISE}
The VDIF data from the VLBI backend were searched using a {\tt Heimdall} and FETCH pipeline, while the PSRIX pulsar data were searched with a {\tt PRESTO} pipeline. Details of this analysis can be found in \citet{kirsten_2021_natas,kirsten_2022_natur}. No additional bursts were found in these data beyond the $5$ reported in \citet{kirsten_2022_natur} and \citet{nimmo_2022_natas}, down to a fluence limit of 0.05\,Jy\,ms (for a 7\,$\sigma$, 100\,$\upmu$s duration burst).

\subsection{PSRIX baseband-mode}
The PSRIX filterbank data were searched for single pulses using tools in the {\tt PRESTO} software suite, matching the analysis of the PRECISE PSRIX data in Section \,\ref{sec:PRECISE}. No bursts were found down to a fluence limit of 0.05\,Jy\,ms (for a 7\,$\sigma$, 100\,$\upmu$s duration burst). 

\section{Burst analysis}
\label{sec:analysis}
\subsection{Dispersion measure}
\label{sec:dm}

\begin{figure*}
\centering
        {\includegraphics[height=90mm,trim=1.35cm 0cm 1cm 0cm, clip=true]{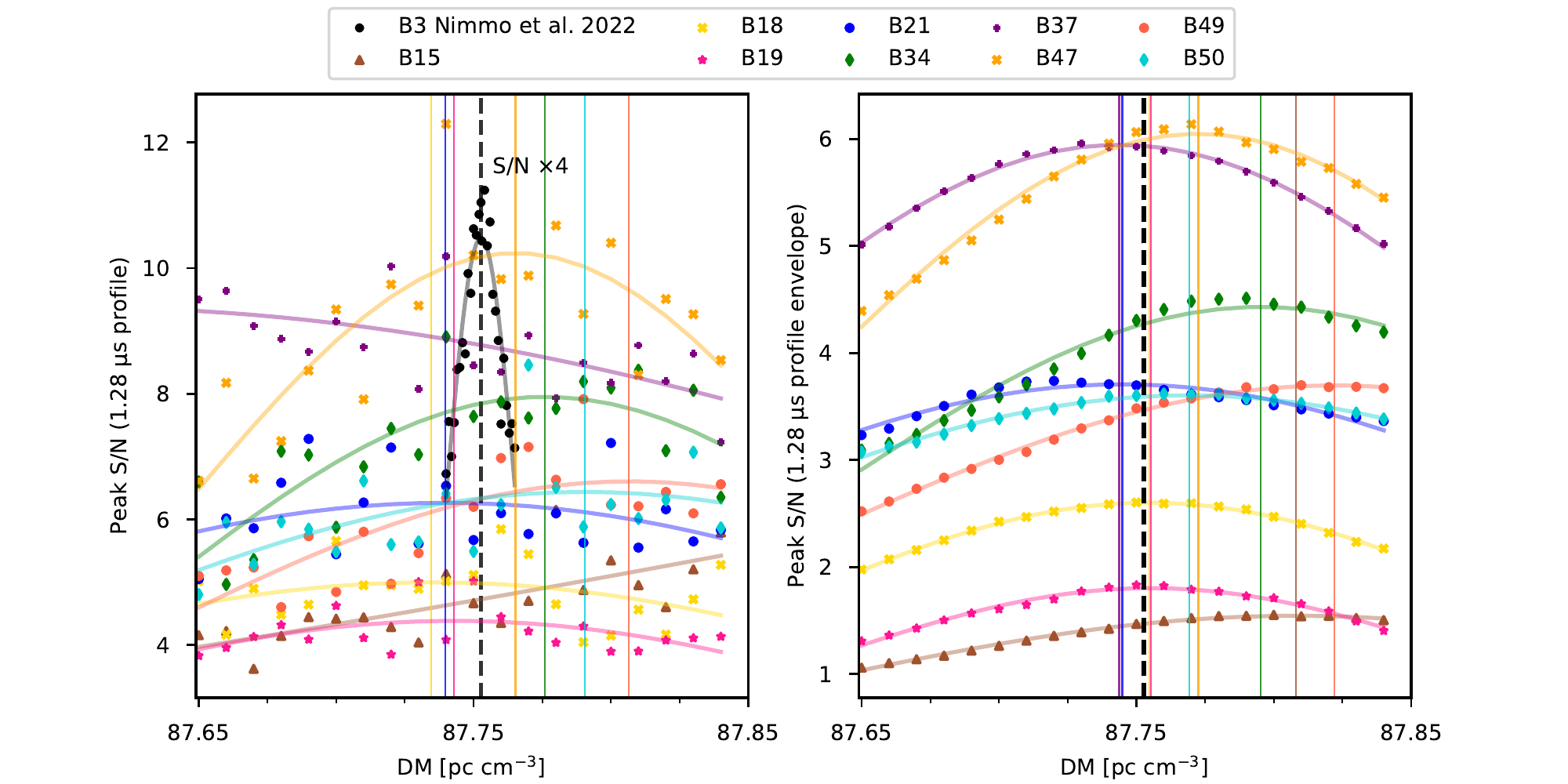}}
  \caption{Peak S/N of 1.28\,$\upmu$s burst profile (left) and peak S/N of the smoothed burst envelope (right) as a function of DM. The bursts were selected based on their S/N values and fine structure observed in the 5.12\,$\upmu$s profiles. Gaussian fits are overplotted for each burst, matching the colour of the burst markers, and the mean of each Gaussian fit are shown by the coloured vertical lines. For comparison, a similar analysis of burst B3 from \citet{nimmo_2022_natas} is overplotted in black (left panel). The best DM measurement from \citet{nimmo_2022_natas}, is shown by the black dashed line in both panels (DM$=87.7527$\,pc\,cm$^{-3}$).}
     \label{fig:dm}
\end{figure*}

Radio waves interact with free electrons on their journey to Earth, causing the lower frequencies to be delayed with respect to the higher frequencies (this relationship is quadratic in frequency; see \citet{lorimer_2004_hpa}). Due to the complex morphology of FRB signals in time and frequency, constraining the quadratic sweep of dispersion can be challenging \citep{hessels_2019_apjl}. Conversely, without accurately correcting for dispersion, structure in the burst profiles could be unresolved. Using the sharp, relatively broadband, temporal features in one burst from \frb\ (on timescales of microseconds), \citet{nimmo_2022_natas} constrained the DM to $87.7527\pm0.0003$\,pc\,cm$^{-3}$, which was sufficient to resolve sub-microsecond timescales. This measurement is $>9\sigma$ lower than the previous measurement from \citet{bhardwaj_2021_apjl}, which could be due to temporal evolution, spectral evolution, or unresolved burst structure in the \citet{bhardwaj_2021_apjl} results. While it is possible that this DM difference reflects changes in the local environment of \frb, it is also possible that it is the result of a turbulent interstellar medium (ISM) in the Milky Way, as observed for pulsars (e.g. \citealt{hobbs_2004_mnras}), or in M81. We note that the DM measurements have not accounted for Doppler shifting from Earth's orbital motion and rotation, although this should likely only contribute a $\sim 0.009$\,pc\,cm$^{-3}$ difference in DM measurements between our observations and the previous DM measurement.

We selected bursts in our sample that have both a high S/N, and sharp structure in the 5.12\,$\upmu$s profile (Figure\,\ref{fig:fullburst}): namely B18, B19, B21, B34, B37, B47, B49 and B50 from 2022 January 14, as well as the narrowest burst in our sample (B15 from 2022 January 14, despite it having a relatively low S/N). Using {\tt digifil}, we created 32-bit, coherently dedispersed (DM$=$87.7527\,pc\,cm$^{-3}$) total intensity filterbank data containing the selected bursts, with 1.28\,$\upmu$s time resolution and 0.7813\,MHz frequency resolution. To measure the DM, we incoherently shifted the frequency channels using DMs in the range 87.65--87.85\,pc\,cm$^{-3}$ in steps of 0.01\,pc\,cm$^{-3}$. In Figure\,\ref{fig:dm} we plot the peak S/N of the 1.28\,$\upmu$s profile as a function of DM. In contrast to the results of \citet{nimmo_2022_natas}, there is no clear micro-structure observed in these bursts, meaning the S/N does not rise and fall with DM as sharply. Note that the peak S/N versus DM mean of burst B36 is $<87.65$\,pc\,cm$^{-3}$ (the lower limit of the x-axis scale in Figure\,\ref{fig:dm}), and we confirm the burst is visibly undercorrected at that `best-fit' DM. We also plot the peak of the burst envelope as a function of DM, created by smoothing the 1.28\,$\upmu$s profile using a low-pass filter. The envelopes more clearly rise and fall with DM. In the absence of burst structure on microsecond (or shorter) timescales, we used the burst envelopes to measure a S/N weighted average DM of $87.77\pm0.05$\,\,pc\,cm$^{-3}$, which is in agreement with the results of \citet{nimmo_2022_natas}. We therefore conclude that the DM of \frb\ has not changed by more than 0.15\,pc\,cm$^{-3}$ ($3$-$\sigma$) over the $\sim$\,1\,year period of observation, and proceed using a DM of 87.7527\,pc\,cm$^{-3}$ \citep{nimmo_2022_natas} for the analysis of the burst sample presented in this work. As mentioned in Section\,\ref{sec:1us}, this uncertainty in the DM results in a maximum smearing within frequency channels less than the time resolution for both 512\,channel (1.28\,$\upmu$s) and 2048\,channel (5.12\,$\upmu$s) data products created in this work.

\subsection{Burst characterisation}
\label{sec:burstchar}
For each burst, we coherently dedispersed and channelised the DADA baseband data to 2048\,channels (0.1953\,MHz and 5.12\,$\upmu$s frequency and time resolution, respectively) using {\tt digifil}. This 32-bit Stokes I data is used to determine the burst properties for all bursts, with the exception of burst B2 on 2022 January 14, and both B1 and B2 on 2022 February 21, where only the 40.96\,$\upmu$s/0.1953\,MHz pulsar data were retained. Frequency channels contaminated by RFI were masked manually for each burst. The data were downsampled before measuring their properties: a summary of the resolutions used for the analysis can be found in Table\,\ref{tab:burst_properties}. The frequency channels are shifted to correct for dispersion, and normalised such that the mean and standard deviation of the noise in each channel is 0 and 1, respectively. 

\begin{figure*}
\centering
        {\includegraphics[height=80mm,trim=1.4cm 0cm 2cm 0.1cm, clip=true]{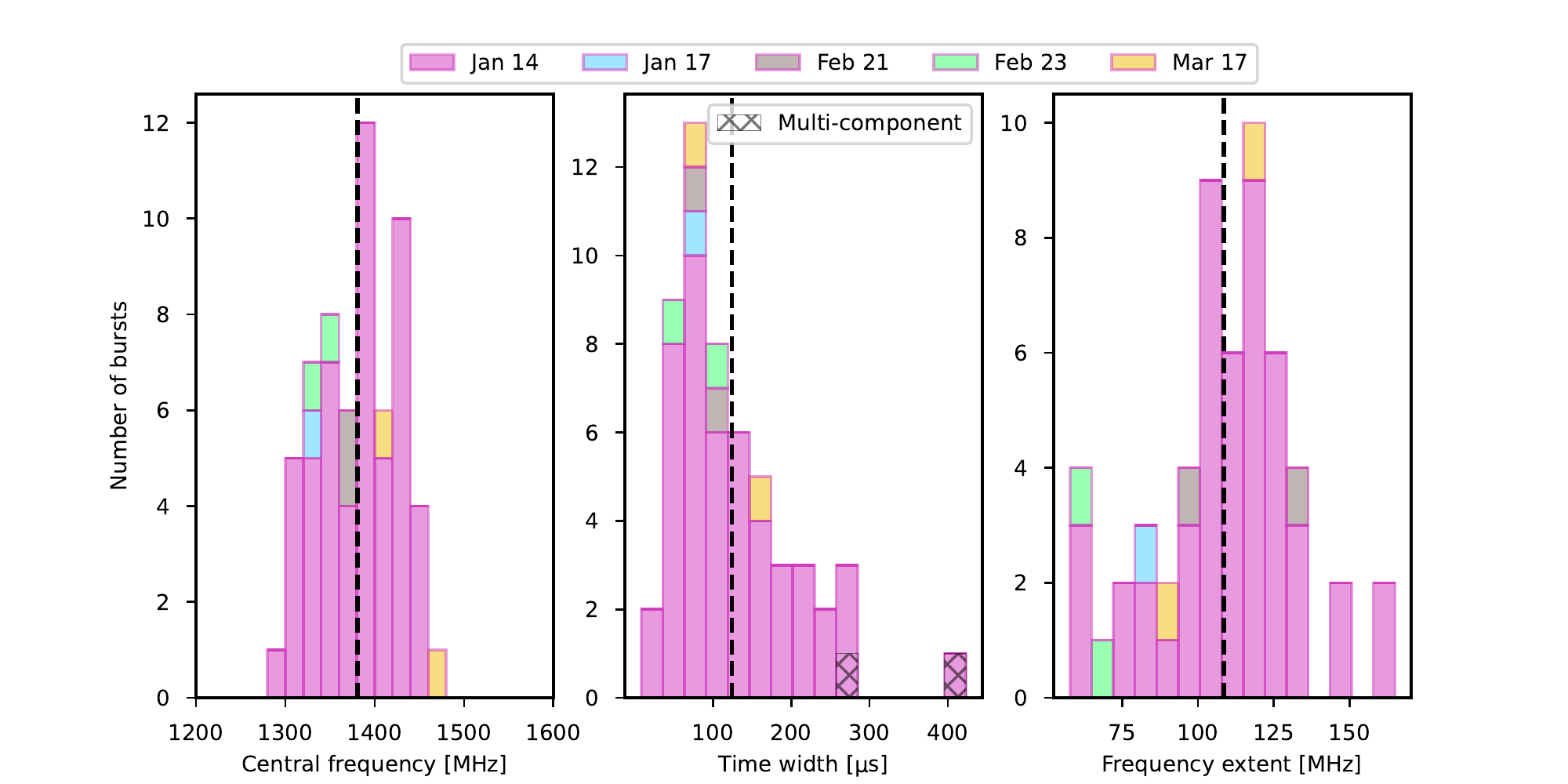}}
  \caption{Histograms of the burst central frequencies (left panel), temporal widths (middle panel) and frequency extents (right panel). The colours represent the different observing epochs and match the colours used throughout this work (see, e.g., Figure\,\ref{fig:fullburst}). In the middle panel, the hatched bins represent the two bursts where we have clearly identified multiple distinct components (B13 and B51 on 2022 January 14). The longer temporal widths are therefore a reflection of the sub-burst separation, while the individual burst components are in fact more narrow. The black dashed line represents the mean value. }
     \label{fig:hist}
\end{figure*}

To measure the burst extent in time and frequency we computed the 2-dimensional autocorrelation function (ACF) of the dynamic spectra. We fitted a 2-dimensional Gaussian to the ACF, from which we derived the burst width and frequency extent reported in Table\,\ref{tab:burst_properties}. In a few cases (marked in Table\,\ref{tab:burst_properties}), the low S/N and strong scintillation structure resulted in visibly under-estimated burst extents from the ACF analysis. For these bursts, we do not report burst widths and frequency extents, and instead assume the average values from the other bursts at that observing epoch, to use for fluence calculations. For all bursts, we fitted a 2-dimensional Gaussian to the dynamic spectrum to determine the centroid of the burst in time and frequency. The coloured bars in Figures\,\ref{fig:somebursts} and \ref{fig:fullburst} highlight the 1-$\sigma$ (dark) and 2-$\sigma$ (light) burst extents, Time\,$=0$ represents the burst centroid in time, which we define as the time of arrival (ToA) of the burst, and the horizontal coloured line on the burst spectrum represents the frequency centroid. Note for B13 and B51 on 2022 January 14, we observe $2$ clear burst components. The ToA and frequency centroid are, therefore, the centre of the means of a Gaussian fit to each component. In Figure\,\ref{fig:hist} we show histograms of the central frequencies, temporal widths and frequency extents of the bursts, with mean values of 1.38\,GHz, 129\,$\upmu$s and $111$\,MHz, respectively. 

\begin{figure}
\centering
        {\includegraphics[height=80mm,trim=0cm 0.2cm 0cm 1.5cm, clip=true]{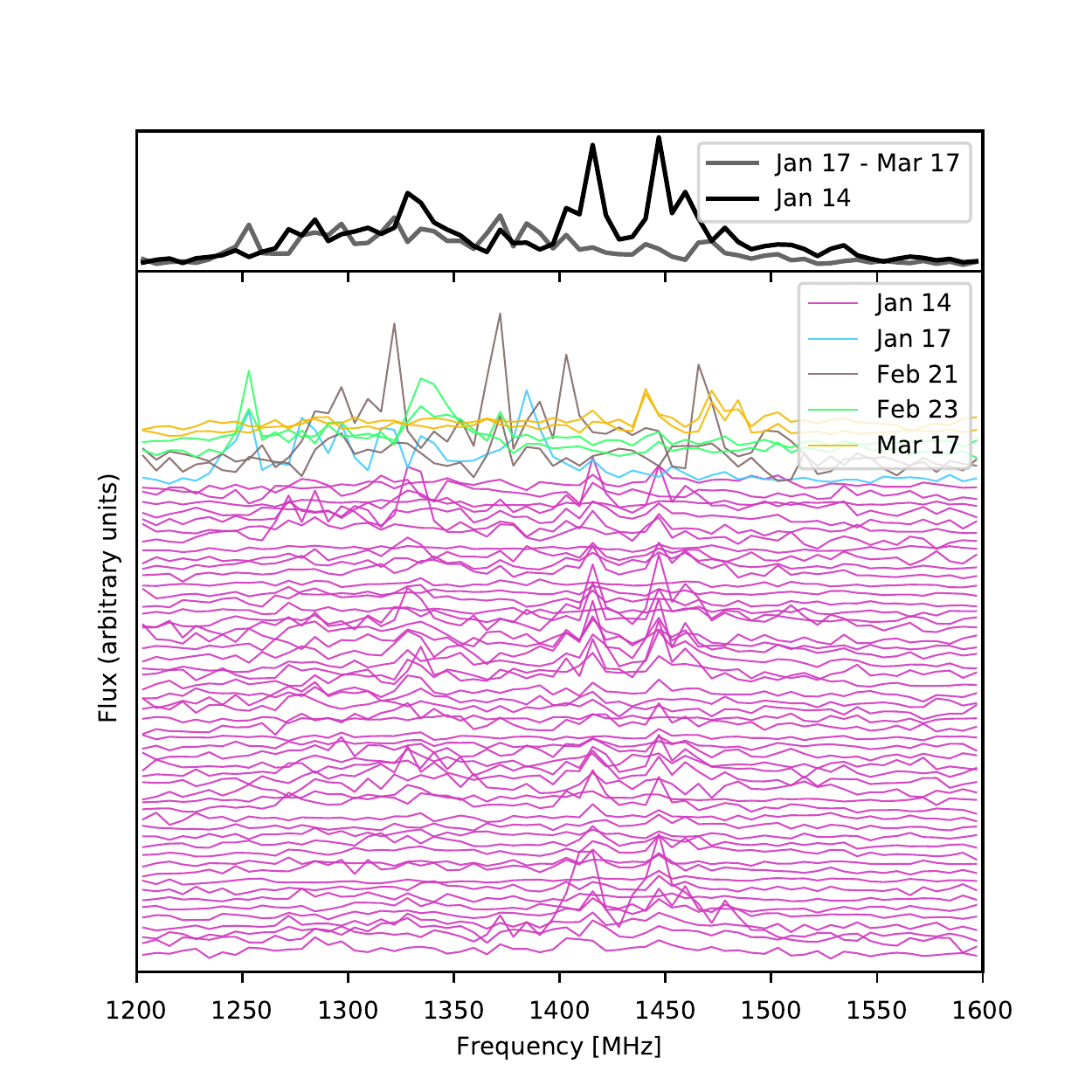}}
  \caption{Burst spectra at 6.25\,MHz resolution. As shown in the legend, the colours represent bursts detected on different days. Strong single-bin spectral features due to scintillation are apparent and persistent through individual observing epochs. The average burst spectra (weighted by burst S/N) for 2022 January 14 bursts (black) and for bursts detected on the other days (grey) are shown in the top panel. In addition to the narrow-band spectral structure due to scintillation, there is also spectral structure on the scale of $\sim100$\,MHz, consistent with the measured burst frequency extents  (Figure\,\ref{fig:hist}). On 2022 January 14, two $\sim100$\,MHz bumps are visible in the stacked spectrum at $\sim1300$ and $\sim1450$\,MHz, reminiscent of the spectral structure seen in the single burst presented in \citet{majid_2021_apjl}. Note the flattening of the burst spectra at $\sim 1550$\,MHz arises from frequent flagging of data in that spectral range due to persistent RFI.}
     \label{fig:spectra}
\end{figure}

The central frequencies are potentially influenced by scintillation from the Milky Way's ISM since \frb\ bursts exhibit strong scintillation spectral structure (visible in their dynamic spectra; see Figure\,\ref{fig:somebursts}). The strong scintles can skew the 2-dimensional Gaussian fit, used to determine the burst centroid. In addition, the central frequencies of the $2$ bursts detected on each of February 21, 23 and March 17, fall to the same side of the mean value in the distribution during a single epoch. The influence of scintillation on the central frequency must be minimal, however, since the bursts on January 17, February 21, 23 and March 17 agree well with the January 14 distribution (Figure\,\ref{fig:hist}). The expected frequency scale arising due to ISM scintillation is $\sim3$\,MHz \citep{cordes_2002_arxiv}, consistent with the single-bin spectral structure observed in the 6.25\,MHz-resolution spectra (Figure\,\ref{fig:spectra}), that persists through individual observations. Since the measured burst spectral extents are $\sim100$\,MHz, a factor of $\sim30$ higher than the scintillation scale, the burst frequency extents are likely not heavily influenced by scintillation from the Milky Way interstellar medium.

The burst ToAs (Table\,\ref{tab:burst_properties}) are corrected to the Solar System Barycentre at infinite frequency (using a dispersion constant of 1/($2.41\times 10^{-4}$)\,MHz$^{2}$\,pc$^{-1}$\,cm$^{3}$\,s, and the VLBI position of \frb; \citealt{kirsten_2022_natur}). The burst profiles are converted to S/N units by subtracting the mean of local noise data (containing no signal), and then dividing by the standard deviation of the noise. The burst profiles are then converted to physical units (Jy) using the radiometer equation \citep{cordes_2003_apj}, and typical values for Effelsberg's system temperature (20\,K) and gain (1.54\,Jy\,K$^{-1}$). These values are uncertain at the 20\% level, which dominates the errors on the flux density. We also add a 3\,K contribution from the cosmic microwave background \citep{mather_1994_apj}, and a sky background temperature of 0.8\,K, which is derived by extrapolating from the 408\,MHz sky map \citep{remazeilles_2015_mnras}, using a spectral index of $-2.7$ \citep{reich_1988_aa}. The fluence is then calculated by summing over the $\pm2\sigma$ burst extent in time {\it and} frequency. Using the known distance to \frb\ (3.63\,Mpc; \citealt{kirsten_2022_natur}), we computed the isotropic-equivalent spectral luminosity of each burst and report these values alongside the burst fluences in Table\,\ref{tab:burst_properties}. 

There are no clear trends of the burst properties with time through the 2022 January 14 observation (Figure\,\ref{fig:props_time}). Perhaps the burst widths and fluences are slightly decreasing through the observation, 
\onecolumn
\begin{longtable}{lccrccccc}
\caption{\label{tab:burst_properties}Burst properties.}\\
\endfirsthead
\caption* {\textbf{Continued}: Burst properties.}\\\toprule
\endhead
\endfoot
\bottomrule
\endlastfoot
\hline
\hline
{Burst}& {Time of Arrival$^{\mathrm{a}}$} & {Fluence$^{\mathrm{b}}$} & {S/N$^{\mathrm{c}}$} & {Spectral Luminosity} & {Width$^{\mathrm{d}}$} & {Frequency Extent$^{\mathrm{d}}$}  & {Time/Frequency} \\ 
{} & {} & {} & {} & {} & {} & {}  & {Resolution}\\ 
{} & {[MJD]} & {[Jy ms]} & & {[10$^{28}$\,erg\,s$^{-1}$\,Hz$^{-1}$]} & {[$\upmu$s]} & {[MHz]}  & {[$\upmu$s / MHz]}\\ 
\hline  
\multicolumn{4}{l}{2022\,January\,14} \\ 
B1 & 59593.70900072 & 0.13\,$\pm$\,0.03 & 11.7 & 0.50\,$\pm$\,0.10 & 171\,$\pm$\,2 & 188.1\,$\pm$\,1.6  & 5.12 / 12.5 \\ 
B2 & 59593.71119049 & 0.54\,$\pm$\,0.11 & 33.1 & 1.04\,$\pm$\,0.21 & 346\,$\pm$\,2 & 176.0\,$\pm$\,0.8  & 40.96 / 6.2 \\ 
B3 & 59593.72152561 & 0.14\,$\pm$\,0.03 & 12.6 & 0.64\,$\pm$\,0.13 & 152\,$\pm$\,2 & 166.8\,$\pm$\,2.1  & 5.12 / 12.5 \\ 
B4 & 59593.72163241 & 0.08\,$\pm$\,0.02 & 10.8 & 0.81\,$\pm$\,0.16 & 133\,$\pm$\,4 & 138.9\,$\pm$\,1.9  & 40.96 / 6.2 \\ 
B5 & 59593.72285434 & 0.36\,$\pm$\,0.07 & 23.2 & 0.80\,$\pm$\,0.16 & 302\,$\pm$\,1 & 166.5\,$\pm$\,0.5  & 5.12 / 6.2 \\ 
B6 & 59593.72347595 & 0.09\,$\pm$\,0.02 & 6.4 & 0.28\,$\pm$\,0.06 & 207\,$\pm$\,7 & 168.2\,$\pm$\,5.0  & 20.48 / 12.5 \\ 
B7 & 59593.72366265 & 0.11\,$\pm$\,0.02 & 9.2 & 0.36\,$\pm$\,0.07 & 240\,$\pm$\,8 & 245.5\,$\pm$\,7.4  & 40.96 / 12.5 \\ 
B8 & 59593.72377264 & 0.10\,$\pm$\,0.02 & 9.7 & 0.44\,$\pm$\,0.09 & 161\,$\pm$\,4 & 213.1\,$\pm$\,4.5  & 10.24 / 12.5 \\ 
B9 & 59593.72393352 & 0.15\,$\pm$\,0.03 & 7.8 & 0.22\,$\pm$\,0.04 & 445\,$\pm$\,7 & 169.6\,$\pm$\,2.3  & 20.48 / 12.5 \\ 
B10 & 59593.72402411 & 0.26\,$\pm$\,0.05 & 19.7 & 0.71\,$\pm$\,0.14 & 245\,$\pm$\,1 & 192.5\,$\pm$\,0.7  & 5.12 / 6.2 \\ 
B11 & 59593.72406203 & 0.10\,$\pm$\,0.02 & 6.2 & 0.18\,$\pm$\,0.04 & 384\,$\pm$\,10 & 241.7\,$\pm$\,5.9  & 20.48 / 12.5 \\ 
B12 & 59593.72564746 & 0.22\,$\pm$\,0.04 & 14.7 & 0.46\,$\pm$\,0.09 & 322\,$\pm$\,2 & 192.7\,$\pm$\,0.8  & 5.12 / 6.2 \\ 
B13 & 59593.72597616 & 0.32\,$\pm$\,0.06 & 17.8 & 0.49\,$\pm$\,0.10 & 431\,$\pm$\,3 & 195.0\,$\pm$\,1.1  & 5.12 / 6.2 \\ 
B14 & 59593.72658502 & 0.17\,$\pm$\,0.03 & 10.5 & 0.35\,$\pm$\,0.07 & 328\,$\pm$\,5 & 195.9\,$\pm$\,2.5  & 10.24 / 12.5 \\ 
B15 & 59593.72694393 & 0.05\,$\pm$\,0.01 & 12.2 & 2.29\,$\pm$\,0.46 & 14\,$\pm$\,1 & 99.8\,$\pm$\,6.7  & 1.28 / 6.2 \\
B16 & 59593.72696785 & 0.08\,$\pm$\,0.02 & 7.0 & 0.28\,$\pm$\,0.06 & 192\,$\pm$\,6 & 212.9\,$\pm$\,6.5  & 10.24 / 6.2 \\ 
B17 & 59593.72767015 & 0.32\,$\pm$\,0.06 & 26.2 & 1.03\,$\pm$\,0.21 & 211\,$\pm$\,1 & 177.0\,$\pm$\,0.5  & 5.12 / 6.2 \\ 
B18 & 59593.72777136 & 0.16\,$\pm$\,0.03 & 12.6 & 0.50\,$\pm$\,0.10 & 217\,$\pm$\,2 & 185.8\,$\pm$\,1.5  & 5.12 / 6.2 \\ 
B19 & 59593.72778012 & 0.06\,$\pm$\,0.01 & 8.9 & 0.71\,$\pm$\,0.14 & 55\,$\pm$\,2 & 171.5\,$\pm$\,5.6  & 5.12 / 6.2 \\ 
B20 & 59593.72835081 & 0.21\,$\pm$\,0.04 & 20.9 & 0.99\,$\pm$\,0.20 & 140\,$\pm$\,1 & 187.9\,$\pm$\,0.8  & 5.12 / 6.2 \\ 
B21 & 59593.72836174 & 0.18\,$\pm$\,0.04 & 19.9 & 1.01\,$\pm$\,0.20 & 124\,$\pm$\,1 & 180.3\,$\pm$\,1.0  & 5.12 / 6.2 \\ 
B22 & 59593.72876785 & 0.06\,$\pm$\,0.01 & 8.5 & 0.59\,$\pm$\,0.12 & 99\,$\pm$\,4 & 167.1\,$\pm$\,5.6  & 20.48 / 6.2 \\ 
B23 & 59593.72956599 & 0.08\,$\pm$\,0.02 & 10.7 & 0.68\,$\pm$\,0.14 & 84\,$\pm$\,2 & 205.6\,$\pm$\,3.8  & 5.12 / 6.2 \\ 
B24 & 59593.72966465 & 0.11\,$\pm$\,0.02 & 8.3 & 0.30\,$\pm$\,0.06 & 251\,$\pm$\,4 & 218.5\,$\pm$\,3.0  & 10.24 / 6.2 \\ 
B25 & 59593.72978254 & 0.11\,$\pm$\,0.02 & 7.2 & 0.22\,$\pm$\,0.04 & 366\,$\pm$\,12 & 270.3\,$\pm$\,9.0  & 20.48 / 12.5 \\ 
B26 & 59593.73021930 & 0.17\,$\pm$\,0.03 & 16.8 & 0.87\,$\pm$\,0.17 & 133\,$\pm$\,1 & 177.7\,$\pm$\,1.0  & 5.12 / 6.2 \\ 
B27 & 59593.73065954 & 0.09\,$\pm$\,0.02 & 7.4 & 0.29\,$\pm$\,0.06 & 215\,$\pm$\,9 & 224.1\,$\pm$\,8.7  & 20.48 / 12.5 \\ 
B28 & 59593.73098736 & 0.08\,$\pm$\,0.02 & 9.5 & 0.65\,$\pm$\,0.13 & 89\,$\pm$\,3 & 171.5\,$\pm$\,3.0  & 5.12 / 6.2 \\ 
B29 & 59593.73118289 & 0.12\,$\pm$\,0.02 & 9.7 & 0.74\,$\pm$\,0.15 & 111\,$\pm$\,3 & 96.2\,$\pm$\,2.5  & 20.48 / 6.2 \\ 
B30$^{\mathrm{e}}$ & 59593.73118737 & 0.12\,$\pm$\,0.02 & 9.4 & 0.39\,$\pm$\,0.08 & -- & --  & 40.96 / 25.0 \\ 
B31 & 59593.73127253 & 0.27\,$\pm$\,0.05 & 27.4 & 1.49\,$\pm$\,0.30 & 126\,$\pm$\,1 & 146.5\,$\pm$\,0.5  & 5.12 / 6.2 \\ 
B32 & 59593.73139847 & 0.07\,$\pm$\,0.01 & 9.8 & 0.64\,$\pm$\,0.13 & 76\,$\pm$\,4 & 193.5\,$\pm$\,8.0  & 10.24 / 12.5 \\
B33 & 59593.73162219 & 0.45\,$\pm$\,0.09 & 30.4 & 1.19\,$\pm$\,0.24 & 249\,$\pm$\,1 & 137.3\,$\pm$\,0.4  & 5.12 / 6.2 \\ 
B34 & 59593.73176535 & 0.16\,$\pm$\,0.03 & 20.9 & 1.38\,$\pm$\,0.28 & 80\,$\pm$\,1 & 168.0\,$\pm$\,1.1  & 5.12 / 6.2 \\ 
B35 & 59593.73215265 & 0.24\,$\pm$\,0.05 & 20.8 & 0.86\,$\pm$\,0.17 & 187\,$\pm$\,1 & 187.4\,$\pm$\,0.7  & 5.12 / 6.2 \\ 
B36$^{\mathrm{e}}$ & 59593.73228158 & 0.09\,$\pm$\,0.02 & 5.9 & 0.27\,$\pm$\,0.05 & -- & -- & 10.24 / 25.0 \\ 
B37 & 59593.73230289 & 0.19\,$\pm$\,0.04 & 27.4 & 1.86\,$\pm$\,0.37 & 71\,$\pm$\,1 & 185.4\,$\pm$\,0.8  & 5.12 / 6.2 \\ 
B38$^{\mathrm{e}}$ & 59593.73244190 & 0.11\,$\pm$\,0.02 & 9.4 & 0.37\,$\pm$\,0.07 & -- & -- & 20.48 / 12.5 \\ 
B39 & 59593.73245115 & 0.11\,$\pm$\,0.02 & 6.7 & 0.26\,$\pm$\,0.05 & 280\,$\pm$\,8 & 192.1\,$\pm$\,5.5  & 10.24 / 12.5 \\ 
B40 & 59593.73278832 & 0.17\,$\pm$\,0.03 & 13.2 & 0.84\,$\pm$\,0.17 & 139\,$\pm$\,2 & 101.0\,$\pm$\,0.0  & 5.12 / 6.2 \\ 
B41 & 59593.73324614 & 0.04\,$\pm$\,0.01 & 2.5 & 0.07\,$\pm$\,0.01 & 405\,$\pm$\,1 & 214.8\,$\pm$\,3.9  & 20.48 / 12.5 \\ 
B42 & 59593.73325516 & 0.08\,$\pm$\,0.02 & 4.7 & 0.13\,$\pm$\,0.03 & 380\,$\pm$\,9 & 215.7\,$\pm$\,4.7  & 10.24 / 6.2 \\ 
B43 & 59593.73364935 & 0.27\,$\pm$\,0.05 & 23.3 & 0.95\,$\pm$\,0.19 & 192\,$\pm$\,1 & 199.9\,$\pm$\,0.6  & 5.12 / 6.2 \\ 
B44 & 59593.73369348 & 0.19\,$\pm$\,0.04 & 14.6 & 0.54\,$\pm$\,0.11 & 231\,$\pm$\,2 & 200.6\,$\pm$\,1.1  & 5.12 / 6.2 \\ 
B45$^{\mathrm{e}}$ & 59593.73403931 & 0.11\,$\pm$\,0.02 & 8.0 & 0.35\,$\pm$\,0.07 & -- & -- & 20.48 / 25.0 \\ 
B46 & 59593.73420280 & 0.07\,$\pm$\,0.01 & 3.9 & 0.11\,$\pm$\,0.02 & 468\,$\pm$\,14 & 203.5\,$\pm$\,6.0  & 10.24 / 12.5 \\ 
B47 & 59593.73421902 & 0.24\,$\pm$\,0.05 & 27.7 & 2.33\,$\pm$\,0.47 & 69\,$\pm$\,1 & 126.7\,$\pm$\,0.6  & 5.12 / 6.2 \\ 
B48 & 59593.73505336 & 0.10\,$\pm$\,0.02 & 9.8 & 0.50\,$\pm$\,0.10 & 135\,$\pm$\,3 & 206.4\,$\pm$\,3.5  & 5.12 / 12.5 \\ 
B49 & 59593.73520064 & 0.13\,$\pm$\,0.03 & 15.1 & 0.80\,$\pm$\,0.16 & 111\,$\pm$\,1 & 193.6\,$\pm$\,1.4  & 5.12 / 6.2 \\ 
B50 & 59593.73588107 & 0.13\,$\pm$\,0.03 & 17.1 & 1.03\,$\pm$\,0.21 & 92\,$\pm$\,1 & 178.6\,$\pm$\,1.2  & 5.12 / 6.2 \\ 
B51 & 59593.73632506 & 0.17\,$\pm$\,0.03 & 7.5 & 0.16\,$\pm$\,0.03 & 704\,$\pm$\,20 & 274.7\,$\pm$\,7.3  & 20.48 / 6.2 \\ 
B52$^{\mathrm{e}}$ & 59593.73726150 & 0.09\,$\pm$\,0.02 & 6.9 & 0.29\,$\pm$\,0.06 & -- & --  & 10.24 / 12.5 \\ 
B53 & 59593.73726998 & 0.11\,$\pm$\,0.02 & 9.3 & 0.51\,$\pm$\,0.10 & 150\,$\pm$\,2 & 131.8\,$\pm$\,2.5  & 10.24 / 12.5 \\ 
\hline 
\multicolumn{4}{l}{2022\,January\,17} \\ 
B1 & 59596.29769138 & 0.56\,$\pm$\,0.11 & 49.9 & 3.06\,$\pm$\,0.61 & 127\,$\pm$\,1 & 134.6\,$\pm$\,0.3  & 5.12 / 6.2 \\ 
\hline 
\multicolumn{4}{l}{2022\,February\,21} \\ 
B1 & {59631.91114554} & 0.10\,$\pm$\,0.02 & 10.2 & 0.49\,$\pm$\,0.10 & 196\,$\pm$\,6 & 216.2\,$\pm$\,5.0 & 40.96 / 12.5 \\ 
B2 & {59631.91878154} & 0.32\,$\pm$\,0.06 & 30.3 & 1.54\,$\pm$\,0.31 & 150\,$\pm$\,0 & 160.7\,$\pm$\,1.0 & 40.96 / 6.2 \\ 
\hline 
\multicolumn{4}{l}{2022\,February\,23} \\ 
B1 & 59633.63710622 & 0.21\,$\pm$\,0.04 & 24.6 & 2.36\,$\pm$\,0.47 & 63\,$\pm$\,0 & 102.7\,$\pm$\,0.6 & 5.12 / 6.2 \\ 
B2 & 59633.64288987 & 0.27\,$\pm$\,0.05 & 20.9 & 1.10\,$\pm$\,0.22 & 163\,$\pm$\,0 & 119.0\,$\pm$\,0.5 & 5.12 / 6.2 \\ 
\hline 
\multicolumn{4}{l}{2022\,March\,17} \\ 
B1 & 59655.95339151 & 0.35\,$\pm$\,0.07 & 26.4 & 0.95\,$\pm$\,0.19 & 245\,$\pm$\,0 & 199.6\,$\pm$\,0.6  & 5.12 / 6.2 \\ 
B2 & 59655.95391696 & 0.10\,$\pm$\,0.02 & 11.0 & 0.67\,$\pm$\,0.13 & 115\,$\pm$\,0 & 148.8\,$\pm$\,6.3  & 10.24 / 12.5 \\ 
\hline 
\multicolumn{9}{l}{$^{\mathrm{a}}$ Corrected to the Solar System Barycentre at infinite frequency using a DM of 87.7527\,pc\,cm$^{-3}$ \citep{nimmo_2022_natas},}\\
\multicolumn{9}{l}{\hspace{0.2em} a dispersion constant of  1/($2.41\times 10^{-4}$)\,MHz$^{2}$\,pc$^{-1}$\,cm$^{3}$\,s, and the VLBI \frb\ position \citep{kirsten_2022_natur}.} \\
\multicolumn{9}{l}{\hspace{0.2em} The times quoted are dynamical times (TDB).}\\
\multicolumn{9}{l}{$^{\mathrm{b}}$ Computed within the $\pm2\sigma$ region of the burst temporal width assuming Effelsberg's system temperature and gain is 20\,K}\\
\multicolumn{9}{l}{\hspace{0.2em} and 1.54\,Jy\,K$^{-1}$, respectively, and also considering a cosmic microwave background temperature of 3\,K and }\\
\multicolumn{9}{l}{\hspace{0.2em} additional sky background temperature of 0.8\,K (see Section\,\ref{sec:burstchar}). }\\
\multicolumn{9}{l}{$^{\mathrm{c}}$ Boxcar S/N: defined as the sum of the burst profile in S/N units within $\pm2\sigma$ of the burst temporal width normalised}\\
\multicolumn{9}{l}{\hspace{0.2em} by the $\pm2\sigma$ width in time bins.}\\
\multicolumn{9}{l}{$^{\mathrm{d}}$ Defined as the full-width at half maximum of the Gaussian fit to the ACF divided by $\sqrt{2}$.}\\
\multicolumn{9}{l}{$^{\mathrm{e}}$ Burst extents in time and frequency could not be measured accurately due to low S/N and temporal/spectral structure.}\\ \multicolumn{9}{l}{\hspace{0.2em} We assume the mean values from the remaining January 14 burst sample (132\,$\upmu$s and 113\,MHz) to compute the fluence. }\\
\end{longtable}
\twocolumn
but the scatter on the data points is too large to confirm a downward trend. The bursts on other observing days tend to have lower burst widths and frequency extents than the mean values of the January 14 observation, while in general they show higher fluences than average (Figure\,\ref{fig:props_time}). More observations are required to test whether the burst properties are drawn from different distributions as the burst rate changes. 

\begin{figure}
\centering
        {\includegraphics[height=200mm,trim=2cm 2cm 0cm 3cm, clip=true]{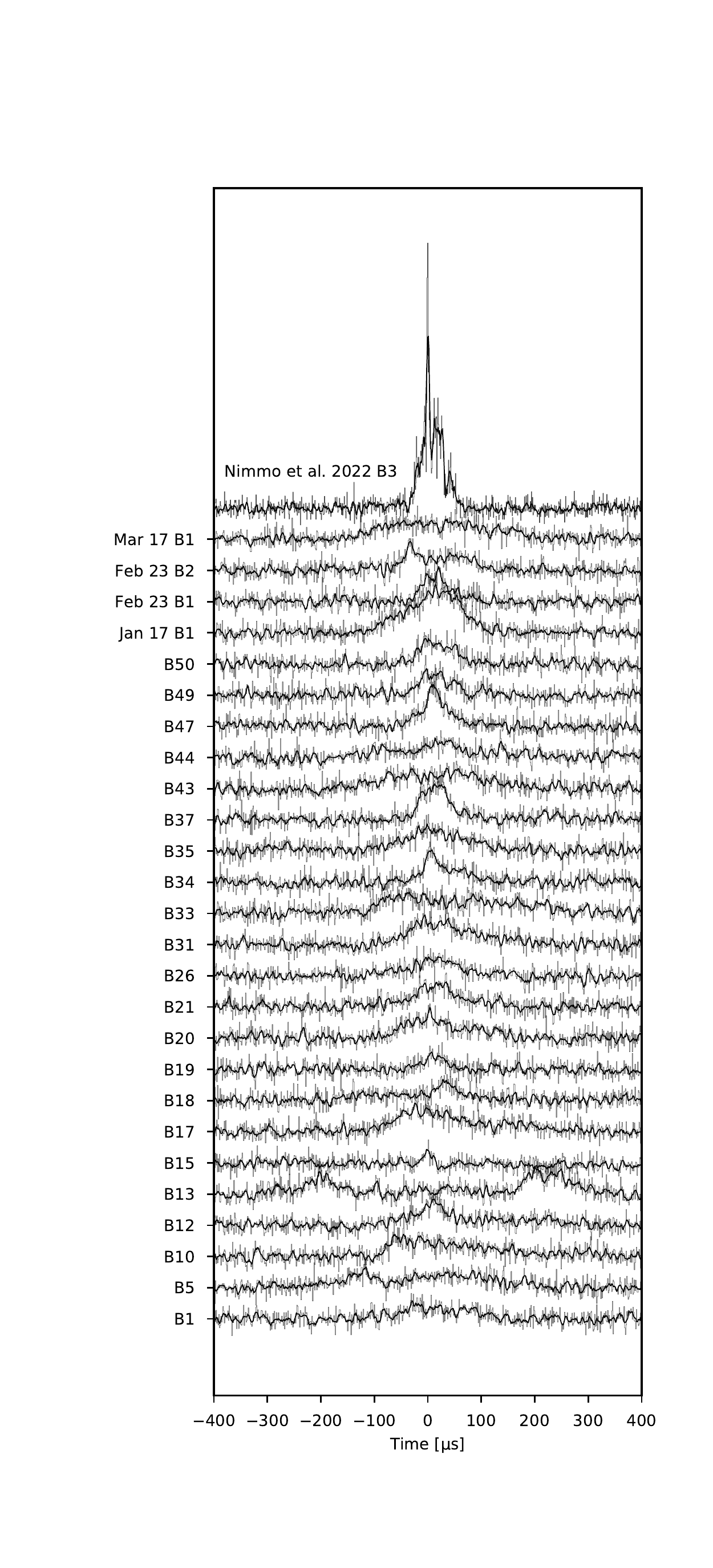}}
  \caption{1.28\,$\upmu$s burst profiles of a sample of high S/N bursts presented in this work (burst names are shown on the y-axis). The burst profiles are created by averaging over the $\pm2\sigma$ burst extent in frequency. The burst profiles are normalised such that they have the same off-burst noise statistics. Overplotted on the 1.28\,$\upmu$s profiles are the smoothed burst profiles, using a Savitzky-Golay filter (\citealt{savitzky_1964_anach}; 11 bins, and polynomial order 2). The top burst profile is B3 from \citet{nimmo_2022_natas} at 1\,$\upmu$s time resolution, which has clear high-S/N microstructure visible. }
     \label{fig:joydiv}
\end{figure}

Motivated by the high S/N microstructure and sub-microsecond structure seen in bursts from \frb\ \citep{nimmo_2022_natas, majid_2021_apjl}, we created higher-time-resolution filterbank data for bursts which have sufficient S/N at 5.12\,$\upmu$s to explore the structure on microsecond timescales. Using {\tt digifil}, we created 32-bit coherently dedispersed total intensity filterbank data containing the bursts with a resolution of 1.28\,$\upmu$s and 0.7813\,MHz in time and frequency, respectively. The burst profiles are created by averaging over the $\pm2\sigma$ burst extent in frequency. In Figure\,\ref{fig:joydiv}, the 1.28\,$\upmu$s burst profiles are shown and compared with the 1\,$\upmu$s profile of burst B3 from \citet{nimmo_2022_natas}, which shows very clear microstructure. We caution that if the DM has changed by $\sim0.1$\,pc\,cm$^{-3}$ ($<3\sigma$ from the DM uncertainty measured in Section\,\ref{sec:dm}), this will result in $\sim60\upmu$s of smearing from 1.5\,GHz to 1.3\,GHz, therefore washing out microsecond structure. It is clear, however, from Figure\,\ref{fig:dm} that the 1.28\,$\upmu$s peak S/N does not rapidly increase in S/N close to $\pm0.1$\,pc\,cm$^{-3}$ around the DM used (B36 is increasing towards 87.65\,pc\,cm$^{-3}$, but as noted in Section\,\ref{sec:dm} is visibly undercorrected at its best-fit value). If the bursts presented in this work had similarly high S/N microstructure as seen in B3 from \citet{nimmo_2022_natas}, this would be evident in the 5.12\,$\upmu$s burst dynamic spectra (Figure\,\ref{fig:fullburst}). We cannot rule out the presence of low S/N microstructure; however, it is evident that the sample of bursts presented in this work do not exhibit the same high-S/N microstructure that has been observed previously from this source.

\begin{figure*}
\centering
{\includegraphics[height=80mm,trim=0cm 0cm 0cm 0cm, clip=true]{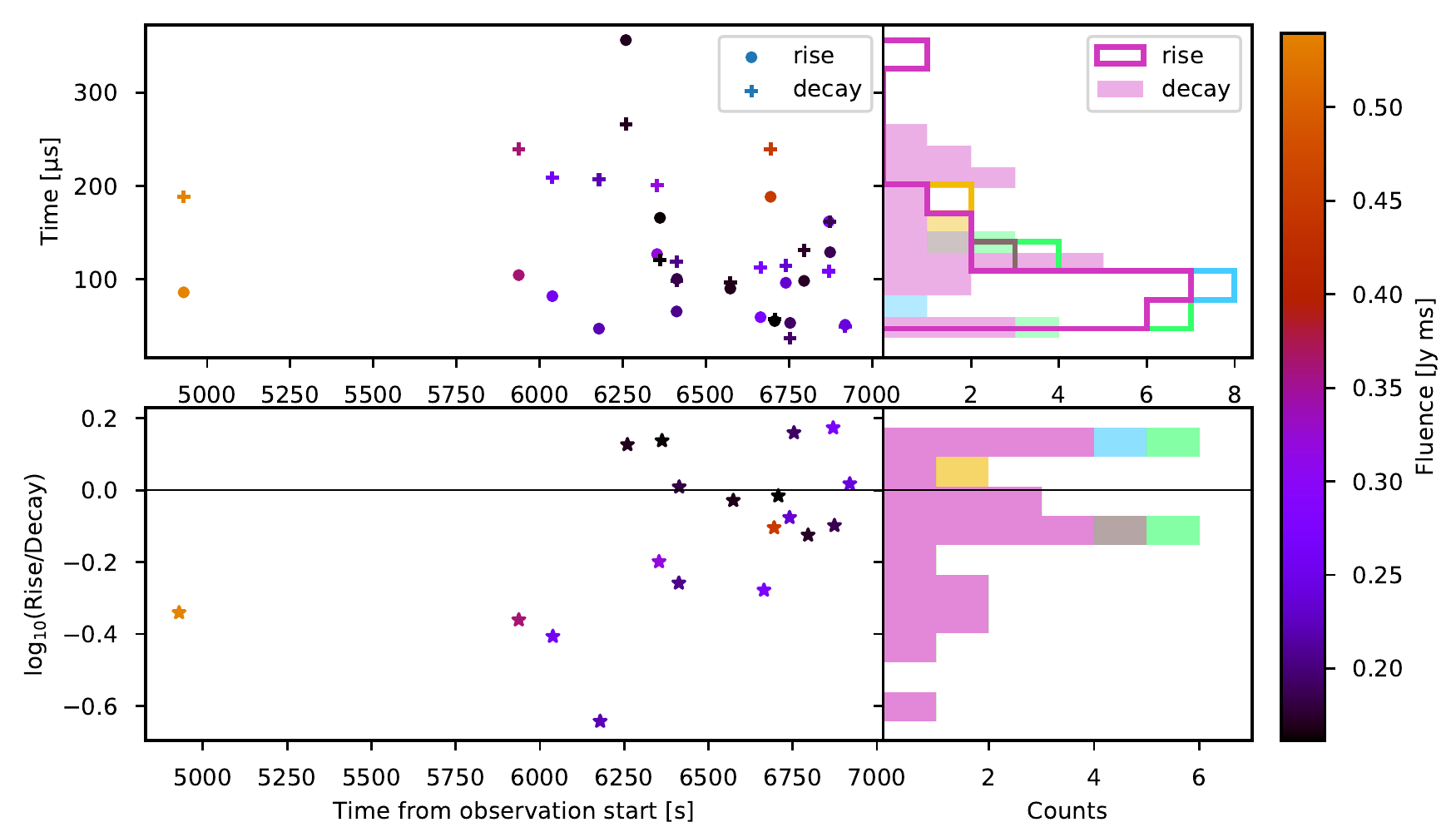}}
  \caption{Rise and decay times of bursts above a fluence threshold of 0.16\,Jy\,ms. Left panels are as a function of time from the beginning of the 2022 January 14 observation, and the colour of the scatter points represent the burst fluence. The right panels are histograms of the rise and decay times (top), and the ratio of rise/decay (bottom), where the colours represent the multiple observing epochs (matching the colour scheme of other plots in this work). The horizontal black line in the bottom plots show the divide between bursts that rise faster than they decay, and vice versa. }
     \label{fig:risedecay}
\end{figure*}

Since this burst sample does not allow us to explore a range of timescales, as in \citet{nimmo_2022_natas}, we instead compute the rise and decay timescales of the high S/N bursts. We use a fluence threshold of $0.16$\,Jy\,ms, which we measure to be the completeness limit for the fluence distribution (see Section\,\ref{sec:energy}). Using this conservative fluence threshold limits the effect of noise on the measurements, while also giving a sufficient sample to study the distribution of rise and decay times (Figure\,\ref{fig:risedecay}). We define the rise time as the time it takes the burst to increase from 10\% to 90\% of the burst energy computed between the peak and peak$-2\sigma_{\rm wid}$, where $\sigma_{\rm wid}$ is the 1\,$\sigma$ burst width. Likewise, the decay time is the time between 90\% to 10\% of the burst energy computed between the peak and peak$+ 2\sigma_{\rm wid}$. We performed this analysis with time resolution of 20.48\,$\upmu$s (with the exception of B2 on January 14 and both bursts on February 21, where only the 40.96\,$\upmu$s pulsar data is available). We find that the rise times are preferentially lower than the decay times (Figure\,\ref{fig:risedecay}). As the fluence decreases, the rise and decay times approach equality, and occasionally  the rise time exceeds the decay time. This evolution with fluence is likely a reflection of the increased influence of noise in the data as the fluence decreases. The range of rise times measured is from 47 to 356\,$\upmu$s, and decay times from 37 to 266\,$\upmu$s. 

\subsection{Burst rates, wait times and clustering}
\begin{figure}
\centering
        {\includegraphics[height=60mm,trim=0cm 0cm 0cm 1cm, clip=true]{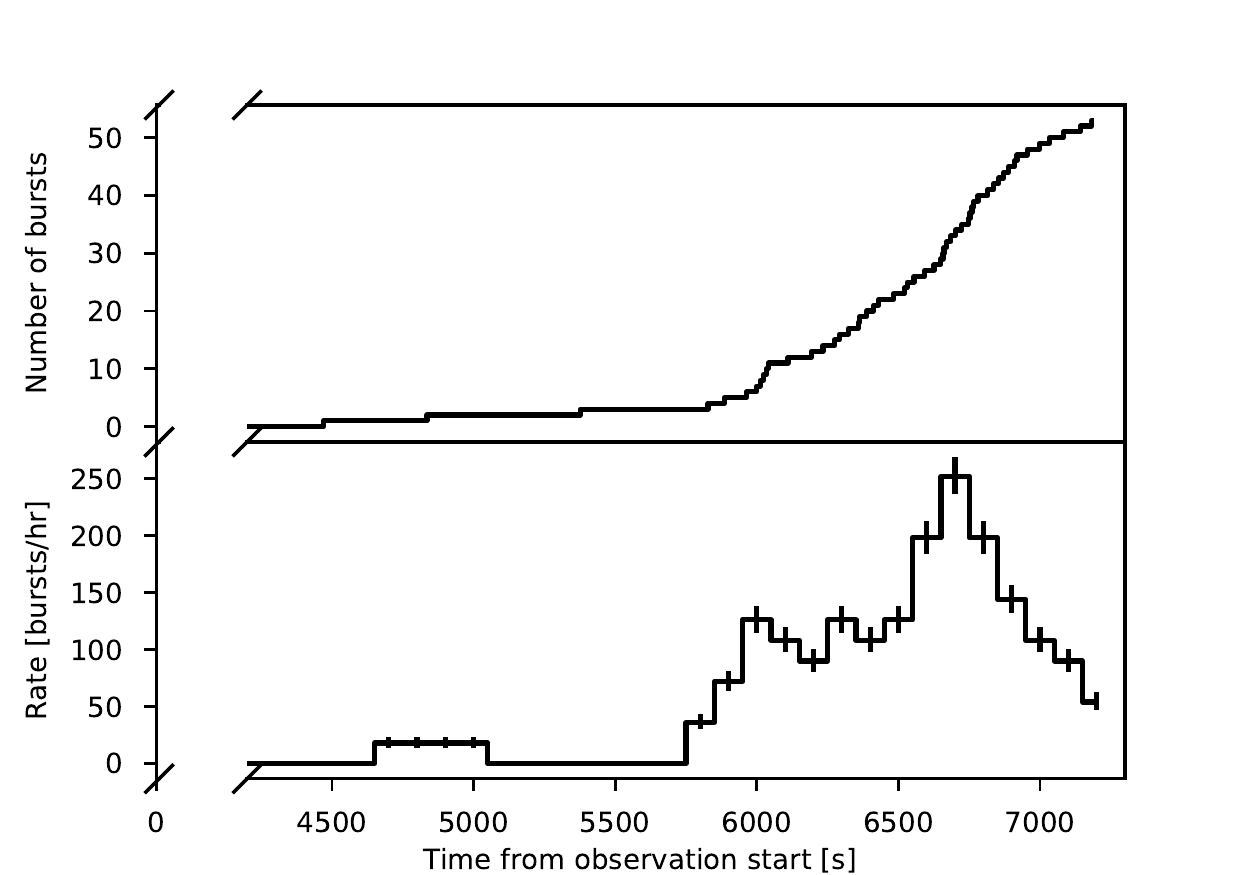}}
  \caption{Cumulative burst distribution (top) and evolution of burst rate per 200\,s time interval (bottom) during the 2022 January 14 \frb\ burst storm. The error bars on the rate are Poissonian. The x-axis is broken to show that there were no bursts detected in the first $\sim$4700\,s of the observation.  }
     \label{fig:rate}
\end{figure}

Reported in Table\,\ref{tab:observations} are the average burst rates per observation during our \frb\ monitoring campaign. For burst rates throughout this paper we report Poisson errors. Even during the observation on 2022 January 14, the burst rate is changing: all of the 53 bursts were discovered in the final $\sim40$\,minutes of a 2-hr observation. We computed the burst rate in 200-s time chunks (Figure\,\ref{fig:rate}) to monitor the evolution of burst rate through the observation. We find that the burst rate ramps up to a maximum of $252^{+17}_{-16}$\,bursts/hr, before falling back down towards the end of the observation. Excluding the first two bursts, which occur significantly earlier than the remaining 50, the burst storm has an approximate duration of 20\,min. We do caution that the storm extends to the end of our observation, beyond which we do not know the activity behaviour of \frb.

\begin{figure*}
\centering
        {\includegraphics[height=100mm,trim=0cm 0cm 0cm 1cm, clip=true]{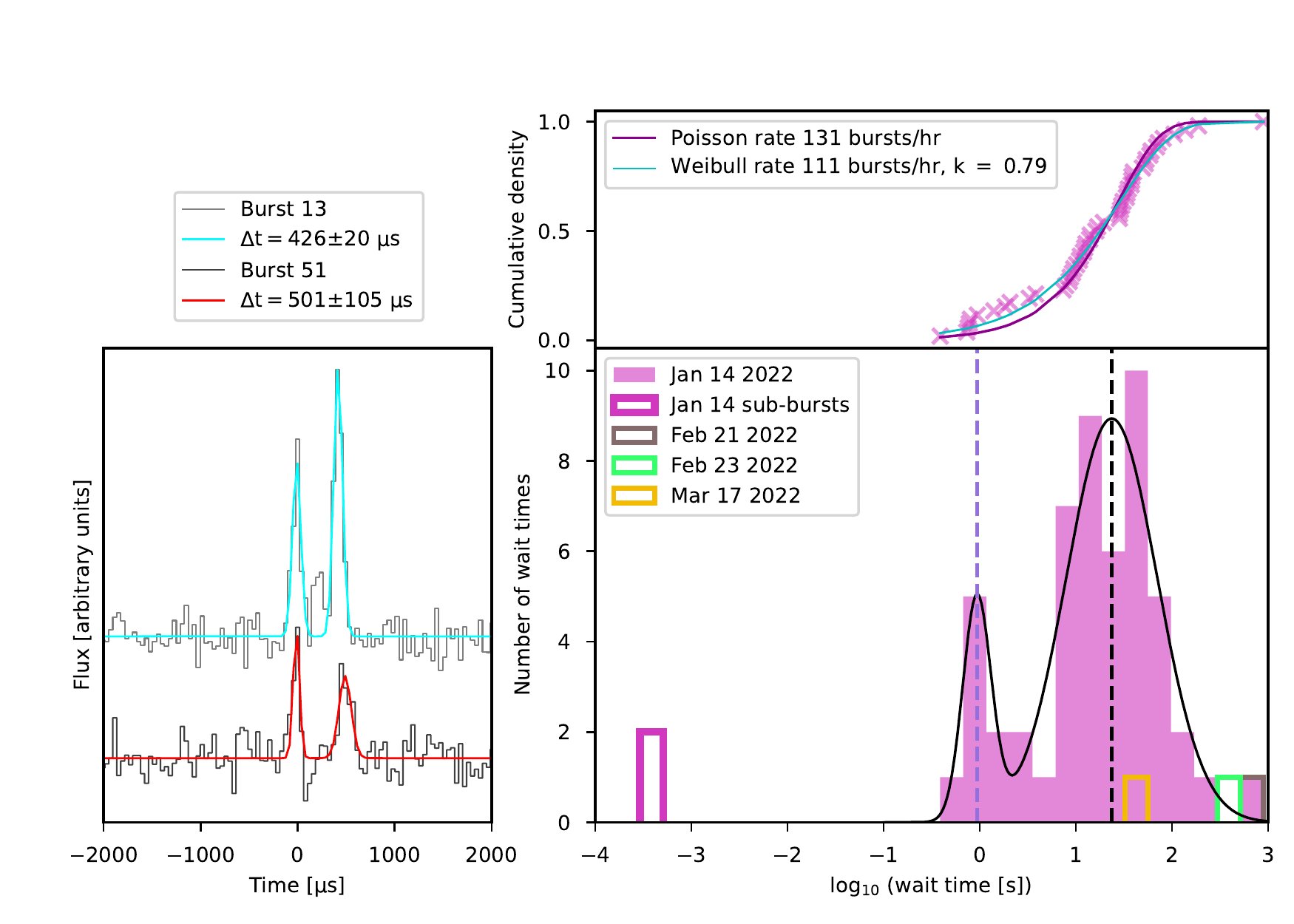}}
  \caption{Wait time distribution (right sub-figure, bottom panel) and cumulative distribution of wait times (right sub-figure, top panel). The purple wait time histogram is that of the bursts on 2022 January 14. We fit the sum of two log-normal distributions to the 2022 January 14 histogram (excluding the sub-burst separations), shown by the black line, and the means of the two log-normals are shown by the purple dashed (0.94$^{+0.07}_{-0.06}$\,s) and black dashed (23.61$^{+3.06}_{-2.71}$\,s) lines. The open grey, green and yellow bins represent the time separation of bursts on 2022 February 21, 23 and March 17, respectively (as highlighted in the figure legend). The burst profiles of the two identified multi-component bursts in our sample, bursts B13 and B51 on 2022 January 14, are shown in the left sub-figure at 40.96\,$\upmu$s time resolution. Overplotted on the profiles are double Gaussian fits to the profiles, and shown in the legend are the measured separations of the Gaussian means. The open purple histogram at short wait times plotted on the wait time distribution represents the measured sub-burst separations. Overplotted on the cumulative wait time distribution (right sub-figure, top panel) are the best-fit Poisson  cumulative density function (CDF) in purple, and the best-fit Weibull CDF in blue.   }
     \label{fig:wait}
\end{figure*}

The distribution of the time difference between consecutive bursts, the so-called `wait times', of the 2022 January 14 burst storm is bi-modal (Figure\,\ref{fig:wait}). We fitted two log-normal functions to the wait-time distribution using least-squares fitting. The best fit log-normal means are 0.94$^{+0.07}_{-0.06}$\,s and 23.61$^{+3.06}_{-2.71}$\,s. We also included bins in the wait time distribution for the time separation of bursts from other observing epochs (Figure\,\ref{fig:wait}), which are all longer than the 23.61$^{+3.06}_{-2.71}$\,s log-normal mean. This is likely a reflection of the highly varying burst rates between observing epochs (Table\,\ref{tab:observations}). The median duration of the $5$ observations with burst detections from our monitoring campaign is $\sim 2$\,hr, meaning we are unable to measure wait times longer than this. The fact that the wait time distribution appears to tail off at $\sim$\,1000\,s is a reflection of being naturally less sensitive to longer wait times due to the limited observation durations. 


Also shown in Figure\,\ref{fig:wait} is the cumulative wait time distribution for the burst storm on 2022 January 14. We test for burst clustering during the burst storm by comparing the cumulative density function (CDF) for a Poisson distribution 
\begin{equation}
    P_{\rm Poisson} = 1 - e^{-t_{\rm wait}R},
\end{equation}
with the CDF for a Weibull distribution
\begin{equation}
    P_{\rm Weibull} = 1 - e^{-(t_{\rm wait}R_{W}\Gamma(1+1/k))^{k}},
\end{equation}
where $t_{\rm wait}$ are the wait times between consecutive bursts, $R$ and $R_{W}$ are the rates for a Poisson and Weibull distribution, respectively, $k$ is the Weibull shape parameter and $\Gamma$ is the gamma function \citep{oppermann_2018_mnras}. The Weibull distribution is equivalent to a Poisson distribution when $k=1$, while $k<1$ implies that the bursts are clustered, with more clustering implied for lower $k$. We performed a least-squares fit of both the Poisson and Weibull distributions to the wait time cumulative distribution. The best-fit Poissonian rate is $131\pm1$\,bursts/hr, and the best-fit Weibull rate is $111\pm3$\,bursts/hr, with shape parameter $0.79\pm0.44$. The reduced $\chi^2$ for the fits are 1.1 (50 degrees of freedom) and 0.2 (49 degrees of freedom) for Poisson and Weibull, respectively, indicating that the burst rate is Poissonian during the burst storm. Furthermore, the large uncertainties on the Weibull shape parameter are also consistent with a Poissonian distribution (the special case of $k=1$). This supports the use of Poissonian error bars on the rate in Figure\,\ref{fig:rate}. 

\begin{figure*}
\centering
        {\includegraphics[height=80mm,trim=0cm 0cm 0cm 0cm, clip=true]{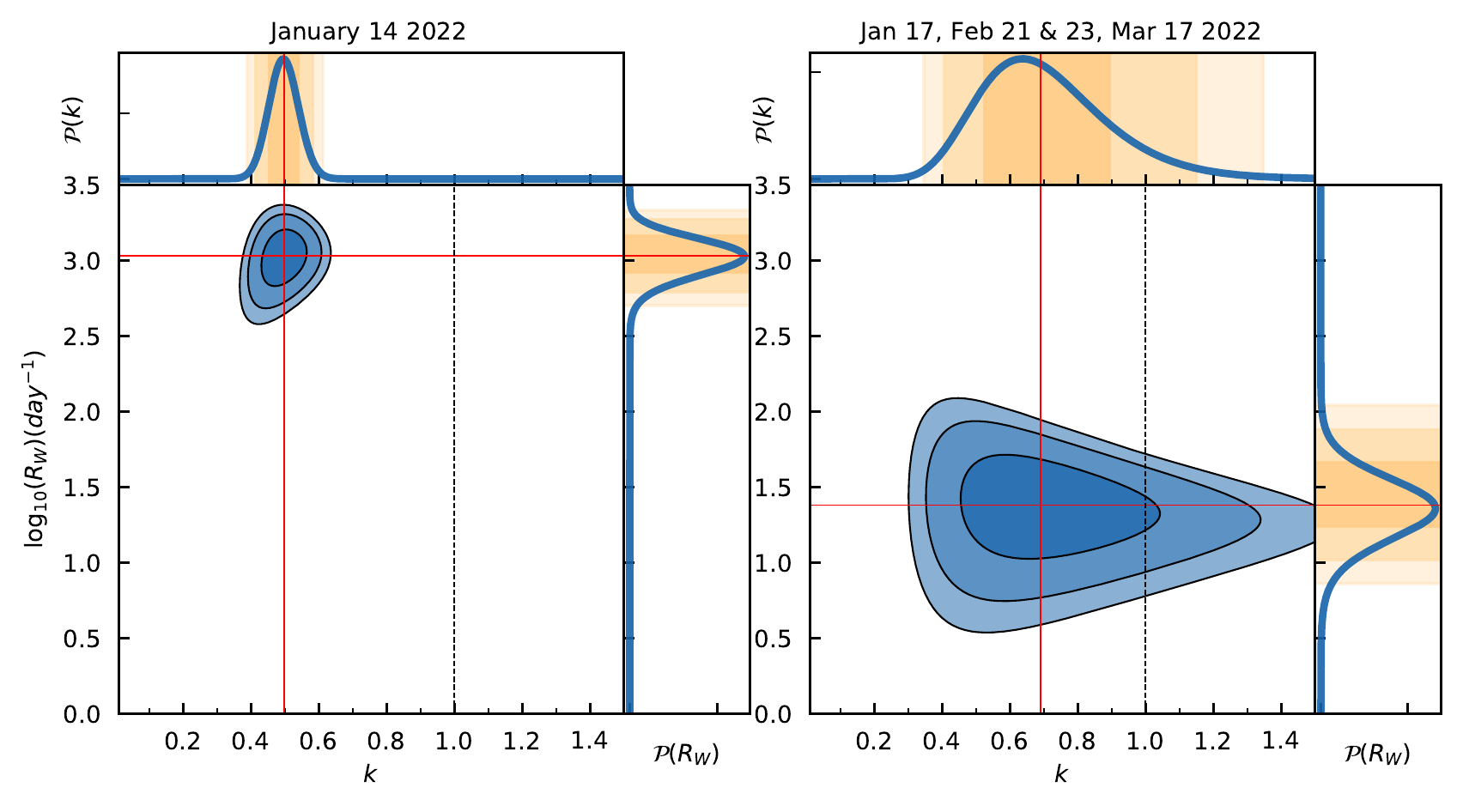}}
  \caption{Posterior distributions of the Weibull shape parameter $k$ and rate $R_{W}$ for the 2022 January 14 observation (left) and the combination of 2022 January 17, February 21, 23 and March 17 observations (right). In each sub-figure the contour plot represents the 2-dimensional distribution, where the contours are 68\%, 95\% and 99\% confidence intervals. The red lines highlight the maximum probability density of the distribution (i.e., the most likely values). The black dashed line highlights $k=1$, which is where the Weibull distribution is equivalent to a Poisson distribution. The top and right panels show the marginal distribution for $k$ and $R_{W}$, respectively. The orange shaded regions represent the 68\%, 95\% and 99\% confidence intervals.  }
     \label{fig:weibull}
\end{figure*}

The fits to the cumulative wait time distribution reflect only the statistics {\it during} the burst storm, and do not account for the $\sim4700$\,s leading up to the burst storm where no bursts were detected (Figure\,\ref{fig:rate}). Following \citet{oppermann_2018_mnras}, and the analysis presented in \citet{kirsten_2021_natas}\footnote{\url{https://github.com/MJastro95/weibull_analysis}}, we calculated the posterior distribution of the 2022 January 14 observation (Figure\,\ref{fig:weibull}, left) and the combined posterior distribution of the observations on January 17, February 21, February 23 and March 17 (Figure\,\ref{fig:weibull}, right). The likelihood function is computed using the burst ToAs reported in Table\,\ref{tab:burst_properties} relative to the beginning of the observation, and incorporates any gaps between the beginning of the observation and the first burst, and the final burst to the end of the scan. We do not include the non-detection observations, since we are specifically interested in exploring whether the difference between 2022 January 14 and other detection days is solely the burst rate. To combine observations we are assuming that the scans are independent and calculate the likelihood of the data as the product of the likelihoods of the individual observations. The prior ($f(k,R_{W})$) is defined as uniform, and the posterior distribution is calculated as 
\begin{equation}
    {\rm Post}(k, R_{W} | d) \propto \mathcal{L}(d | k, R_{W})f(k,R_{W}), 
\end{equation}
for the likelihood of the data $d$, $\mathcal{L}(d | k, R_{W})$. 

We find that for the 2022 January 14 observation, the most likely values of $k$ and $R_{W}$ are $0.50^{+0.04}_{-0.05}$ and $1079^{+377}_{-242}$\,day$^{-1}$, respectively, where the uncertainties reflect the 68\% confidence interval. This indicates that the bursts are highly clustered (since $k<1$), which is evident already from Figure\,\ref{fig:rate}. In contrast, the combined observations of January 17, February 21, 23 and March 17 result in most likely values of $k=0.69^{+0.20}_{-0.17}$ and $R_{W}=24^{+22}_{-7}$\,day$^{-1}$, showing no strong evidence for clustering, and consistent with $k=1$ within the 95\% confidence interval (Figure\,\ref{fig:weibull}). 

\subsection{Periodicity searches}
\label{sec:period}
In this work we have a relatively large burst sample, mostly concentrated in a short time interval on 2022 January 14. We, therefore, used various methods to search for a periodic arrival time of the bursts during the 2022 January 14 observation, as well as a periodicity in the activity of bursts over the $>1$\,yr span of observations. Note that we only include the bursts presented in this work and in \citet{kirsten_2022_natur} for the periodicity analysis since they were all detected with Effelsberg (observations with comparable sensitivity) at the same observing frequency (frequency-dependent activity is seen in FRB~20180916B; \citealt{pleunis_2021_apjl,pastormarazuela_2021_natur}).

\subsubsection{Burst arrival times}
 Using {\tt PRESTO} \citep{ransom_2001_phdt}, we created dedispersed time series (DM$=87.7527$\,pc\,cm$^{-3}$; \citealt{nimmo_2022_natas}) of the entire 2-hr observation, the last $\sim40$\,minutes, and the $\sim15$\,minutes around the peak burst rate of 2022 January 14, after masking the RFI using {\tt rfifind}. The reason for segmenting the data in this manner is to increase sensitivity to the scenario where the periodic emission has `turned on' (e.g., nulling pulsars; \citealt{backer_1970_natur}). The time series were created from the 40.96\,$\upmu$s pulsar backend data since there are roughly a few minutes of missing data from the raw voltages (see Section\,\ref{sec:research}).  These time series were Fourier transformed (using a Fast Fourier Transform, FFT) and then searched for periodic signals using {\tt PRESTO}'s {\tt accelsearch}. We first use a maximum Fourier frequency derivative of $8$\,bins and $8$ maximum number of harmonics, and second a maximum Fourier frequency derivative of $200$\,bins and $16$ maximum number of harmonics. The only periodic candidates above a S/N threshold of $7$ and coherent power of $100$ are confined to a small spectral range, and therefore are attributed to RFI.

 We then performed a brute force search of an integer divisor of the burst ToAs commonly used for period searches of rotating radio transients (RRATs; \citealt{mcLaughlin_2006_natur})\footnote{Using {\tt rrat\_period} in the PRESTO \href{https://github.com/scottransom/presto/blob/master/python/presto/psr\_utils.py}{\tt psr\_utils} package.}. We performed this test twice: once on all bursts excluding the first two which are separated in time from the main outburst of FRBs, and second a subset of those bursts which have a wait time from the previous burst $>2.5$\,s. For the latter, we are using only bursts in the long wait time log-normal (Figure\,\ref{fig:wait}), which we attribute to the burst rate, and exclude bursts in the shorter wait time log-normal which we attribute to a typical `event duration' (see Section\,\ref{sec:discwait}). Folding at the best-fit period from each search returns bursts across all burst phases (Figure\,\ref{fig:period}, bottom sub-figure). Therefore, we conclude that there is no strict period in the arrival times of the bursts during the \frb\ burst storm. We note that this method is insensitive to the arrival of bursts at various rotational phases of the progenitor.

To increase our sensitivity to the situation where the bursts arrive at a wider range of phases, or at multiple distinct rotational phases, we folded the observation using a range of trial periods from 1\,ms to 25\,s. We made step sizes of (1/$n_{\rm bins}$)/(2442.5\,s) in frequency space, where 2442.5\,s is the time separation between the first and last burst on 2022 January 14, and $n_{\rm bins}$ are the number of bins the period is divided into: we choose $25$\,bins which corresponds to a minimum duty cycle of 4\%. In the folded observation, we count the number of burst ToAs that fall into each phase bin, and use the $\chi^2$ statistic to compare with a uniform distribution. During the 2022 January 14 observation we have even exposure to all phase bins, but when considering a longer period in the activity of \frb\ (see Section\,\ref{sec:period_act}), the uneven exposure must be accounted for. This analysis is described in detail in \citet{chime_2020_natur_582}. The reduced $\chi^2$ as a function of period for $n_{\rm bins}=25$ is shown in Figure\,\ref{fig:period}. We computed a significance from the $\chi^2$ survival function, and find $4$ periods with significance $\sim3\sigma$. For those 4 period candidates, we fold the burst ToAs using those periods and find that in all cases the burst ToAs appear across $>90$\% of the phase. In Figure\,\ref{fig:period} we plot reduced $\chi^2$ for a small range of periods around the highest significance candidate (period $P=3.40$\,ms), and also plot the histogram of phases after folding the burst ToAs using this period. Since there are no periods that confine the burst ToAs in phase in a statistically significant way, we conclude that we find no evidence for periodicity in the arrival times.

\subsubsection{Burst activity}
\label{sec:period_act}

The repeating FRB~20180916B has a confirmed $16.33\pm0.12$\,day periodicity in its bursting activity \citep{chime_2020_natur_582,pleunis_2021_apjl}. Additionally, FRB~20121102A has a tentative $\sim160$\,day period in its activity \citep{rajwade_2020_mnras,cruces_2021_mnras}. With the $>1$\,yr span of Effelsberg observations, as well as multi-epoch detections, we aim to search for a similar periodic activity from \frb. It may be that the PRECISE detections of 2021 February--April \citep{kirsten_2022_natur}, are one activity cycle, and the 2022 January--March detections presented in this work are a second cycle, giving a period of $\sim11$\,months, and an activity window of $\sim3$\,months (Figure\,\ref{fig:obs}). In this case, however, having only $2$ cycles, and a deficit of observations from June 2021 to December 2021, means this is impossible to confirm. Perhaps, however, there is a shorter period in the activity ($<11$\,months). We used two methods to search for periods between 1.5\,days and 80\,days: we constrain the lower search limit to $1.5$\,days to minimise the effect of the sidereal day and above $80$\,days there are too few cycles to constrain any period meaningfully with the current data.  

Since our observations of \frb\ are not evenly sampled in time, we created a Lomb-Scargle periodogram \citep{lomb_1976_apss,scargle_1982_apj}, from 1.5\,days to 80\,days in 50000 linear steps. We created a list of observation epochs (using the time stamp of the observations beginning), paired with a list of $1$s and $0$s: $1$ when the observation contained at least 1 burst, and $0$ when the observation resulted in a non-detection. Following \citet{vanderPlas_2018_apjs}, we then subtracted the mean of this time series, since the Lomb-Scargle model assumes the time series is centred around the mean. The periodogram is shown in Figure\,\ref{fig:period} (top left sub-figure, bottom panel), with a $\sim$12.5-day candidate highlighted. Note that Lomb-Scargle periodograms have a complicated window function due to the uneven sampling of the data. We produced this window function by making a Lomb-Scargle periodogram of a time series reflecting our observing cadence: the time series is $1$ for days where we had observations, and $0$ for every other day. We confirm that this $\sim$12.5-day candidate is not present in the window function, and is therefore present in the data itself. To determine the 1\,$\sigma$ significance level, we randomly selected 8 days out of our list of observing epochs as `detection days' and compute the maximum value of the Lomb-Scargle periodogram. We repeated this exercise 1000 times and from the distribution of maximum values compute the 1\,$\sigma$ significance level. The 12.5-day candidate is $<1\sigma$ in significance when following this approach.

Additionally, we fold the $>1$\,yr span of observations using periods between 1.5\,days to 80\,days. We step in frequency by 0.1/(390\,days) -- or (1/(10 bins))/(separation of first detection and last detection) -- and bin the data into $3$, $5$, and $10$ bins. We compare detections per bin with a uniform distribution, similar to the analysis we conducted on the single 2022 January 14 observation described above, calibrating for the uneven exposure per bin \citep{chime_2020_natur_582}.  In Figure\,\ref{fig:period} (top left sub-figure, top panel) we show the reduced $\chi^2$ as a function of period using a binning of $3$ bins per period. The 12.5-day candidate is also evident in this analysis, but with similarly low significance. We bootstrap the significance by randomly selecting 8 detection days from our observing epochs and computing the largest reduced $\chi^2$ value, repeating this exercise 100 times, and from the distribution of maximum reduced $\chi^2$ we compute the 1\,$\sigma$ significance level. 

Although there is again a candidate at a period of 12.5\,days, it is not statistically significant. Continued monitoring of \frb\ at 1.4\,GHz is required to confirm whether \frb's activity is periodic similar to the behaviour of FRB~20180916B \citep{chime_2020_natur_582} and the tentative activity period detected for FRB~20121102A \citep{rajwade_2020_mnras,cruces_2021_mnras}. 

\subsection{Energetics}
\label{sec:energy}
\begin{figure*}
\resizebox{\hsize}{!}
        {\includegraphics[height=90mm,trim=1cm 0cm 1cm 1cm, clip=true]{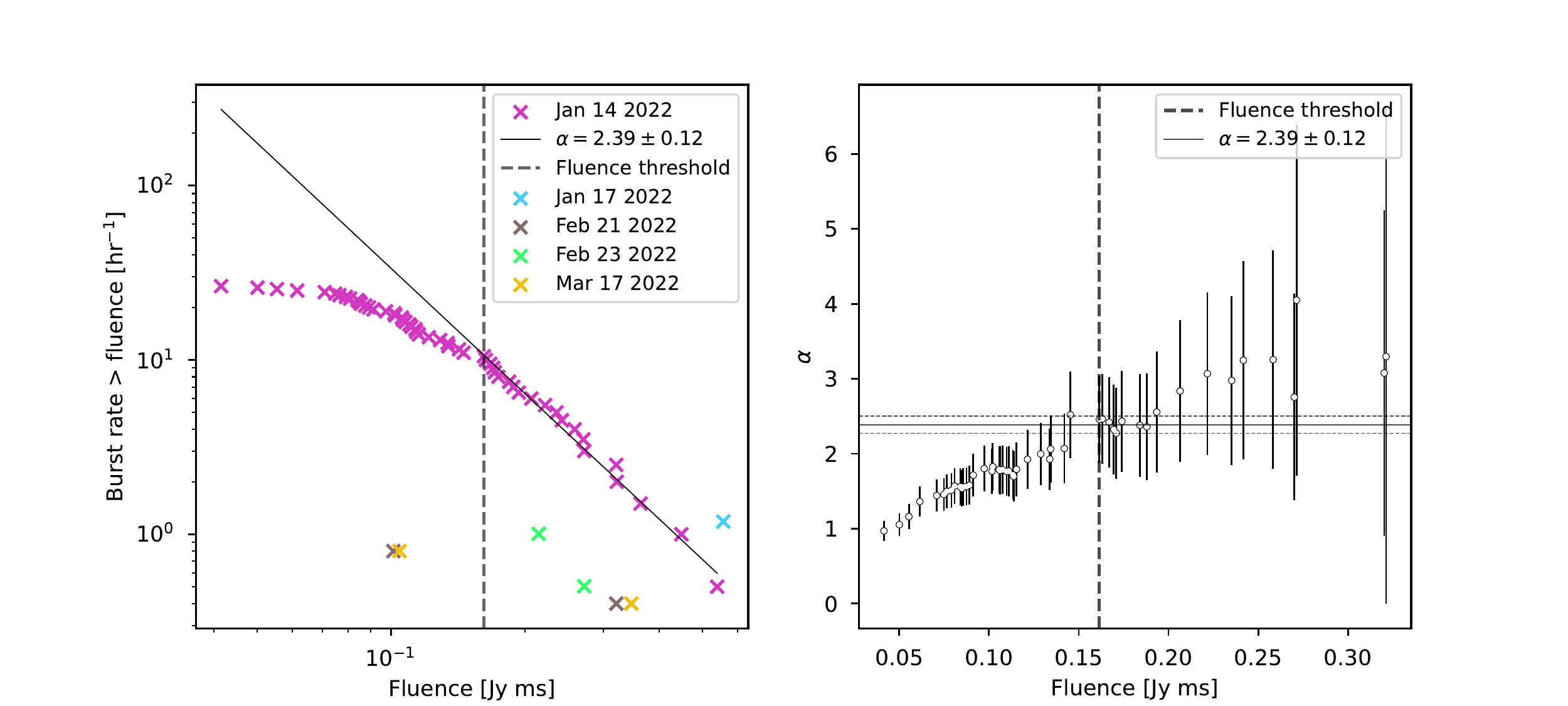}}
  \caption{Left: Distribution of burst fluences. Different coloured markers, as shown in the legend, represent bursts detected on different observing days. The black solid line represents the best fit power-law to the 2022 January 14 bursts above the fluence threshold. Right: The slope of the power-law $\alpha$, determined by Maximum-likelihood estimation \citep{crawford_1970_apj,james_2019_mnras} as a function of fluence. The best fit $\alpha$ from the least-squares fit to the distribution is shown by the horizontal line. In both panels, the dashed vertical line represents the completeness limit of $0.16$\,Jy\,ms.}
     \label{fig:energy}
\end{figure*}

The cumulative distribution of burst fluences appears to turn-over towards low fluences (Figure\,\ref{fig:energy}). This is likely the result of our inability to detect all bursts close to the sensitivity limit of our observations. We must determine a completeness threshold, above which the fluence distribution accurately reflects the source's behaviour. Selecting the completeness threshold, though, is somewhat ambiguous, and can heavily influence the inferred fluence distribution properties. We used the method of maximum-likelihood to measure the optimal fluence threshold resulting in the best power law fit \citep{clauset_2007_arxiv}, under the assumption that the bursts follow a power-law energy distribution above some threshold. This method returns a limit of $0.16$\,Jy\,ms, which is chosen as the completeness limit since it lines up with where the distribution begins to turn-over by eye, and agrees with the limit derived using the radiometer equation \citep{cordes_2003_apj} and assuming a minimum S/N of $10$, a burst width of $300\upmu$s and a burst bandwidth of $150$\,MHz. This threshold is also consistent with where the power law index, $\alpha$ ($R\propto Fl^{-\alpha}$, for burst rate $R$ above some fluence $Fl$), flattens out as a function of fluence threshold (Figure\,\ref{fig:energy}, right), estimated using a Maximum-likelihood method, described in \citet{crawford_1970_apj} and \citet{james_2019_mnras}. 

We fitted the cumulative distribution of burst fluences on 2022 January 14  above a completeness threshold of $0.16$\,Jy\,ms, using a power-law and a least-squares fit method (Figure\,\ref{fig:energy}). The fluences are considered to have 20\% uncertainty, arising due to the uncertainty in the system values for Effelsberg. The best-fit power law has index $\alpha=2.39\pm0.12$. The uncertainties are the quadratic sum of the statistical fit uncertainties, combined with systematic uncertainties derived by sampling 15 fluences above the completeness threshold, performing the same least-squares fit to the distribution and repeating this process 500 times to measure the standard deviation of the power law indices measured.

In Figure\,\ref{fig:energy}, in addition to the fluence distribution of the bursts on 2022 January 14, we also plot the fluences of bursts detected at our other observing epochs. Due to the different burst rates between observations, and having only 1 or 2 bursts per observation outside of the burst storm on 2022 January 14, it is difficult to constrain how the energy distribution changes from epoch-to-epoch, as is seen in other FRBs \citep[e.g.,][]{jahns_2023_mnras}. To test whether the fluences measured on days other than 2022 January 14 are drawn from a different energy distribution, we performed a Kolmogorov-Smirnov (KS) test using the Python {\tt scipy.stats} package tool {\tt ks\_2samp}\footnote{\href{ https://docs.scipy.org/doc/scipy/reference/generated/scipy.stats.ks_2samp.html}{\tt https://docs.scipy.org/doc/scipy/reference/generated/scipy.stats.ks\_2samp.html}}. The critical value for the KS-test is $D=c(p)\sqrt{\left(\frac{N_1+N_2}{N_1N_2}\right)} = 0.66$, where $c(p) = 1.63$ corresponding to a significance level of $0.01$, and $N_1$, $N_2$ are the sizes of the two distributions we are comparing. We find a KS statistic of $0.51 < 0.66$, and therefore we {\it cannot} reject the null hypothesis that the 2022 January 14 fluences and the fluences of the other epochs are drawn from the same distribution. We repeated this exercise adding the first $2$ bursts from 2022 January 14 to the fluences at other epochs, motivated by the fact that these bursts occur $\sim 15$\,minutes before the burst storm, seemingly less related to the high activity (see, e.g., Figure\,\ref{fig:rate}). In this case the critical value is $D=0.59$ and the measured test statistic is $ 0.34 < 0.59$, which still cannot rule out all fluences being drawn from the same distribution. 

\section{Discussion}
\label{sec:discussion}

\frb\ is a singular source: it provides a valuable bridge between the populations of known pulsars and bursting magnetars in the Milky Way and Magellanic clouds, and the much more distant FRBs in extragalactic space \citep{nimmo_2022_natas}. Here we discuss \frb\ in the context of other fast radio transient sources. We show that \frb\ presents many similarities to the phenomena seen from radio-emitting magnetars and repeating FRBs, but its range of luminosities, burst durations, and wait times also distinguish it from these other known sources. It may be possible to reconcile these quantitative differences by invoking an evolutionary sequence, or spectrum of neutron stars, with a range of rotational rates and magnetic field strengths. Alternatively, \frb\ may originate from a qualitatively different source class. 

\subsection{Lack of DM variations}

Comparing our most recent bursts with those detected $>10$ months earlier \citep{nimmo_2022_natas,kirsten_2022_natur}, we constrain DM variations towards \frb\ to $< 0.15$\,pc\,cm$^{-3}$. This is a strong constraint compared to, e.g., the $> 1$\,pc\,cm$^{-3}$ variations seen from FRB~20121102A on timescales of months to years \citep{hessels_2019_apjl,li_2021_natur}. The strong constraint on DM variation is consistent with the conclusion that \frb\ is in a relatively `clean' local environment compared to the extreme magneto-ionic environments of FRB~20121102A \citep{michilli_2018_natur} and FRB~20190520B \citep[e.g.,][]{niu_2022_natur}. Furthermore, this motivates continued searches for \frb\ bursts at low radio frequencies ($< 400$\,MHz) since such detections can measure more subtle variations in the local medium \citep{pleunis_2021_apjl}. Likewise, the lack of DM variation is also consistent with the hypothesis that \frb\ was formed via accretion-induced collapse, or compact binary merger, in its dense globular cluster environment \citep{kirsten_2022_natur,lu_2022_mnras,kremer_2021_apjl}. Note, however, that the Crab pulsar, which is likely the product of a core-collapse supernova, also shows only small DM variations ($\lesssim 0.02$\,pc\,cm$^{-3}$; \citealt{driessen_2019_mnras}). Whereas some repeating FRBs are noisy probes of the intervening magneto-ionised medium, because of their extreme local plasma environments (e.g. \citealt{niu_2022_natur}), \frb\ demonstrates that some repeaters will serve as accurate probes of the intervening magneto-ionised medium.

\subsection{Burst storm \& time dependent burst rate}

The burst rate of \frb\ varies significantly between observing epochs (Table\,\ref{tab:observations}), with a peak rate of $26.5^{+6.2}_{-5.1}$ bursts/hr averaged over the 2022 January 14 observation. The burst rate is also observed to be highly time variable during the 2022 January 14 observation, with no bursts detected until 1.3\,hr into the 2\,hr observation, and the rate (computed in 200\,s time intervals) ramping up to a maximum $252^{+17}_{-16}$\,bursts/hr, before falling back down towards the end of the observation (Figure\,\ref{fig:rate}). Due to this rapid rise and decay in burst rate, we refer to this as the first observed `burst storm' from \frb. 

\begin{figure*}
\centering
        {\includegraphics[height=80mm,trim=0cm 0cm 0cm 0.4cm, clip=true]{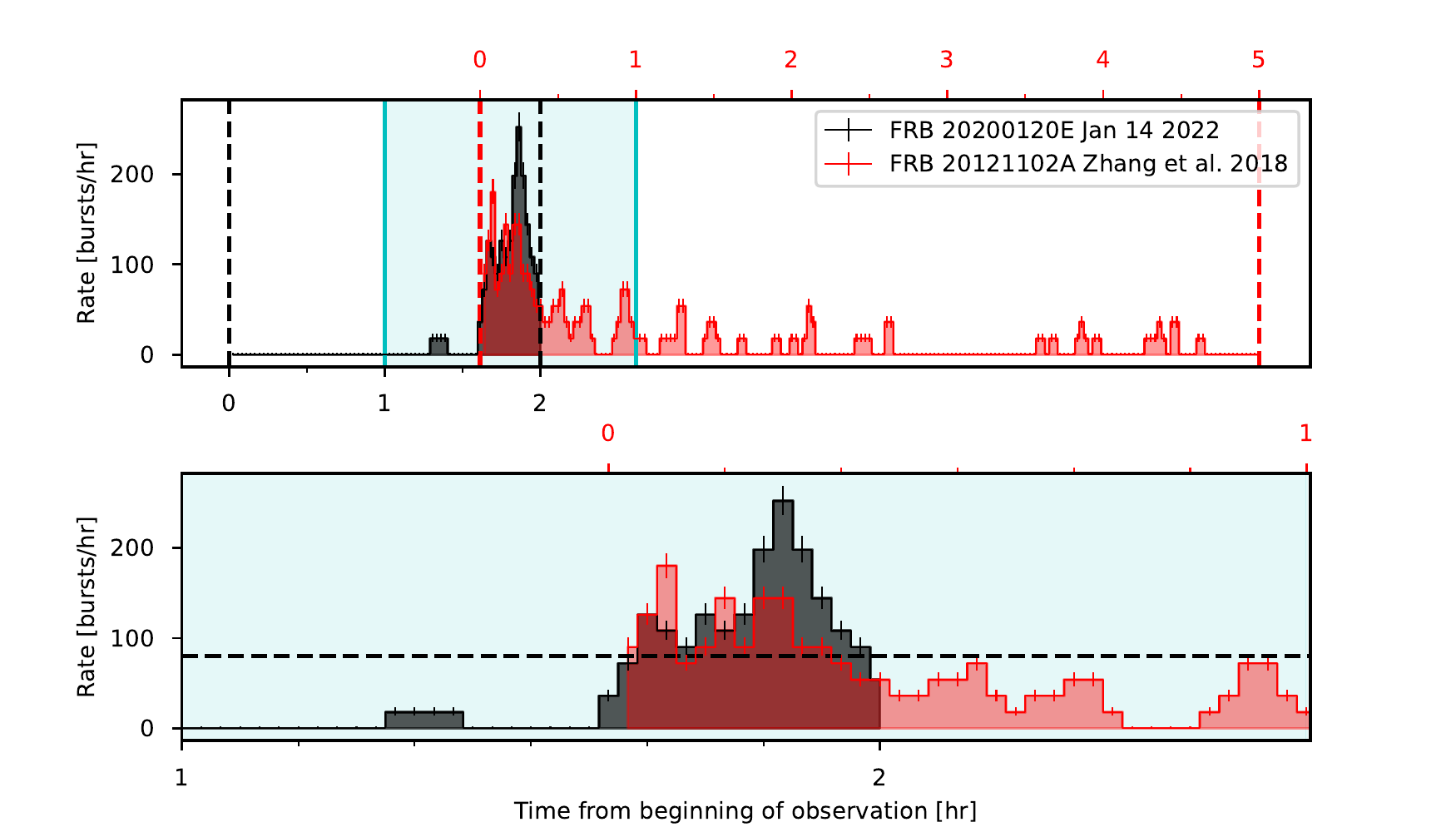}}
  \caption{Comparison between the burst rate as a function of time for the \frb\ burst storm presented in this work, and that of an observation of \rone\ \citep{gajjar_2018_apj,zhang_2018_apjl}. The rates are computed in 200\,s time intervals. For the \frb\ observations the time axis is the lower x-axis (in black) from the beginning of the 2022 January 14 observation, while the upper x-axis (in red) is the time axis for \rone\ from the beginning of the \citet{gajjar_2018_apj} observation. The plots have been arbitrarily aligned in time by eye to highlight the comparable temporal extents. The shaded blue region highlights the extent plotted in the zoom-in panel below. The horizontal dashed line represents a burst rate of 80\,bursts/hr, above which the duration of the storm for both \frb\ and \rone\ are comparable ($\sim$20\,min).}
     \label{fig:compzhang}
\end{figure*}

The first-discovered repeating FRB, FRB~20121102A \citep{spitler_2016_natur}, is one of the few repeating FRBs whose long-term activity has been studied in detail. The burst rate varies significantly between observing epochs \citep{li_2021_natur,hewitt_2022_mnras,jahns_2023_mnras} similar to the behaviour we observe here: the peak rate in $\sim$1-hr observations has been observed to be as high as 218$\pm$16\,bursts/hr, which is less than a factor of 2 higher than the 131$\pm$1\,bursts/hr Poisson rate for \frb\ (Figure\,\ref{fig:wait}). Evidence of burst clustering has been observed for FRB~20121102A \citep{oppermann_2018_mnras, oostrum_2020_aa}. At least some of this burst clustering is related to the apparently periodic activity cycle that FRB~20121102A follows ($\sim$160\,days; \citealt{rajwade_2020_mnras,caleb_2020_mnras}). During these active windows, however, the burst wait times are Poisson distributed, but the rate changes from day-to-day \citep{cruces_2021_mnras,jahns_2023_mnras}. Although there is only a hint of an activity period for \frb, given our observations, we find a similar behaviour: during the burst storm the bursts are consistent with being Poisson distributed (Figure\,\ref{fig:wait}), and the observations on other days with detections (excluding the storm) are consistent with being Poisson distributed as well (Figure\,\ref{fig:weibull}). The observation of 2022 January 14, however, shows clustering on $<1$\,hr timescales (Figures\,\ref{fig:rate},\ref{fig:weibull}). This rapid rise in burst rate, before quickly decreasing in rate has been seen before for FRB~20121102A, although at much higher observing frequencies \citep{gajjar_2018_apj,zhang_2018_apjl}. For comparison, we plot the burst rate as a function of time in $200$\,s intervals for both the \frb\ burst storm, and that of FRB~20121102A (Figure\,\ref{fig:compzhang}). We, unfortunately, do not see the rate drop completely to zero before the end of our observation, and the FRB~20121102A storm had presumably begun before the beginning of the \citet{gajjar_2018_apj} observation. We do, however, find the durations of the two storms, with rate above 80\,bursts/hr, to be comparable ($\sim20$\,minutes). 

Significant change of burst rate over month-to-year timescales has been observed for the highly active repeating FRB~20201124A \citep{lanman_2022_apj}. Since CHIME is a transit telescope, it has almost daily exposure to FRB~20201124A. This provides strong constraints on the burst rate of FRB~20201124A in the years prior to discovery in 2020, and how the burst rate slowly evolved into an outburst in 2021 \citep{lanman_2022_apj}. The burst rate through the outburst varies on day-to-day timescales, rising sharply to the peak reaching a plateau and rapidly turning off \citep{xu_2022_natur}. Perhaps this is a similar phenomenon to what has been observed from \frb\ in this work (Figure\,\ref{fig:obs}). \citet{xu_2022_natur} also observe day-to-day changes in the Weibull $k$ parameter, varying from Poissonian ($k=1$) to clustered ($k<1$) between observations. The peak burst rate per observation at 1.5\,GHz (similar central frequency as the observations presented in this work) is 45.8$^{+7.8}_{-8.3}$\,hr$^{-1}$ \citep{xu_2022_natur}, consistent with our measured burst rate from the 2022 January 14 observation (Table\,\ref{tab:observations}). 

Magnetars are observed to go into outburst, producing tens to hundreds of X-ray bursts per hour \citep{gavriil_2004_apj,israel_2008_apj,vanderhorst_2012_apj}. The Galactic magnetar \sgr, is currently the only known Galactic object that has produced a millisecond-duration radio transient with luminosity comparable to that of the extragalactic FRBs (albeit still 1--2 orders of magnitude weaker than the least luminous FRBs; \citealt{chime_2020_natur_587,bochenek_2020_natur}).  \sgr\ went into outburst in 2020, hours before the FRB-like transient was discovered, with a burst rate of $\sim720$\,bursts/hr \citep{fletcher_2020_gcn,palmer_2020_atel,younes_2020_apjl}. This outburst was observed to have a consistently high rate for at least $20$\,minutes, before rapidly dropping in rate to $\sim29$\,bursts/hr in only 3\,hours. 

In the case of giant pulse emitters, variations in giant pulse rate have been observed between observing epochs: e.g., the Crab pulsar (PSR~B0531+21), where the rate of high fluence ($>130$\,Jy\,ms) giant pulses vary by up to a factor of 5 between observing days \citep{bera_2019_mnras}, and the `Crab twin' PSR~B0540$-$69 in the Large Magellanic Cloud, which has giant pulse rate variations of 65\,/hr to 221\,/hr between epochs separated by $\sim$\,months \citep{geyer_2021_mnras}. 

\subsection{Energetics}
\label{sec:subenerg}
The spectral luminosities of \frb\ bursts are at least two orders of magnitude lower compared to other known repeating FRBs \citep{nimmo_2022_natas}, and even $\sim10\times$ lower than the exceptionally bright event seen from SGR~1935+2154 on April 28 2020 \citep{bochenek_2020_natur,chime_2020_natur_587}. In our unprecedentedly large sample of \frb\ bursts we also see a relatively narrow range of burst fluences, spanning only about an order-of-magnitude from 0.04\,Jy\,ms to 0.6\,Jy\,ms. This limited range is partly due to being strongly sensitivity limited, despite the large aperture of the Effelsberg telescope. Furthermore, many of the bursts we detect are below our nominal completeness threshold of 0.16\,Jy\,ms and only detectable because of their bright, narrow-band scintles. Larger on-sky time ($>100$\,hr) may still reveal \frb\ bursts that are more comparable in their energetics to other repeaters, but nonetheless \frb\ appears to be, at least on average, an anomalously weak source that is only detectable because of its exceptional proximity to Earth \citep{bhardwaj_2021_apjl,kirsten_2022_natur}.

Furthermore, we measure a steep power-law ($\alpha = 2.39\pm0.12$, see Section\,\ref{sec:energy}) burst energy distribution above our fluence threshold of 0.16\,Jy\,ms. Unless \frb\ is found to have a bi-modal and/or time-variable energy distribution, with a flatter tail at high fluences, then it is unlikely that ongoing observing campaigns will detect bursts that are much above a fluence of 2\,Jy\,ms at 1.4\,GHz (corresponding to an isotropic-equivalent spectral luminosity of $\sim 2.4\times10^{29}$\,erg\,s$^{-1}$\,Hz$^{-1}$).
For comparison, \citet{bhardwaj_2021_apjl} find fluences of $\sim2$\,Jy\,ms in the $400-800$\,MHz range. Given that the bursts we detect at 1.4\,GHz are typically $\sim0.2$\,Jy\,ms, this suggests an average spectral index, $S\propto\nu^{-\beta}$ of $\beta\sim2-3$. If this were to continue to low radio frequencies, then the expected average fluence at 150\,MHz is $\sim30-130$\,Jy\,ms, easily detectable by LOFAR or uGMRT. The caveat to this statement is that the higher fluences detected by \citet{bhardwaj_2021_apjl} might reflect the lower sensitivity of CHIME compared with Effelsberg, implying the power law is more shallow than determined here. 

The burst energy distribution of \frb\ is comparable to that of the Crab pulsar \citep{karuppusamy_2010_aa}. It is significantly steeper than the energy distribution seen from FRB~20121102A \citep{li_2021_natur,hewitt_2022_mnras}, but conversely much flatter compared to what has been observed from FRB~20201124A \citep{lanman_2022_apj}. We caution, however, that FRB~20121102A has shown a bi-modal and time-variable burst energy distribution \citep{li_2021_natur,hewitt_2022_mnras}. This bi-modality could indicate that the source produces multiple types of bursts, or that some bursts are apparently boosted in energy due to local propagation effects like plasma lensing \citep{cordes_2017_apj}. Our burst energy distribution for \frb\ is based on a single burst storm and may not be representative of its average behaviour. Future detections of burst storms from \frb\ can test this. The pulse energy distributions of the Crab and other pulsars appear to be stable with time \citep{bera_2019_mnras}. If the burst energy distributions of repeating FRBs are found to be time variable then models of the emission process need to explain this behaviour. 

\subsection{Burst durations and morphology}

The $\sim 100$\,$\upmu$s bursts from \frb\ are on average $>10\times$ shorter-duration compared to other known repeaters \citep{nimmo_2022_natas,pleunis_2021_apj,li_2021_natur,xu_2022_natur}. Previous studies have also detected (sub-)microsecond burst structure from \frb\ \citep{nimmo_2022_natas,majid_2021_apjl}. Coupled with the much lower burst luminosities from \frb\ compared to other repeaters, this suggests that future studies using large burst samples should investigate whether there is a correlation between burst duration and luminosity. Multi-frequency observations of FRB~20121102A have demonstrated that its bursts are on average narrower and less luminous at high radio frequencies \citep{michilli_2018_natur,gajjar_2018_apj,josephy_2019_apjl}. Future observations should aim to establish such a trend for \frb\ as well.

The voltage data we have collected here has also allowed us to constrain how often \frb\ produces ultra-short bursts, on timescales of microseconds or less. The lack of interstellar or intergalactic scattering towards \frb, along with its stable DM, also make it a prime target to explore ultra-short timescales. We find no evidence for microstructure in our sample of \frb\ bursts, nor do we identify any additional bursts in a separate search of the 2022 January 14 data at a time resolution of $1$\,$\upmu$s. While some \frb\ bursts do present structure on (sub-)microsecond timescales \citep{nimmo_2022_natas,majid_2021_apjl}, we conclude that such timescales are relatively rare\footnote{It is, of course, possible that micro-bursts are common but that they typically overlap in time, therefore our results indicate that {\bf isolated} micro-bursts are rare.} and that most bursts have minimum timescales of variation on the order of $>10$\,$\upmu$s.

Nonetheless, we find that the rise times of \frb\ bursts are typically very short: $50-200$\,$\upmu$s. This constrains the size of the emission region to tens of kilometres or less, though relativistic effects may also be relevant. The decay times we measure are typically twice as long, and this asymmetry should be explained in emission models. We note that magnetar X-ray bursts are often also well modelled by a faster rise and slower decay \citep{huppenkothen_2015_apj}.

In any case, the intrinsic asymmetry of the bursts, along with the potential for time-frequency drifts \citep{hessels_2019_apjl}, demonstrates that caution is needed when inferring scattering times from FRBs. Burst B33 from 2022 January 14 is the best example of a `sad trombone' drift in our new burst sample (see Figure~\ref{fig:somebursts}). This effect is also clearly visible in the baseband data from the discovery of \frb\ \citep[][see their Figure~1]{bhardwaj_2021_apjl}, and provides an important phenomenological link with the rest of the known repeater population. In addition, we find that the \frb\ bursts are sometimes narrow band ($\Delta\nu\sim200$\,MHz), as has been seen in other repeaters \citep{hessels_2019_apjl,gourdji_2019_apjl,kumar_2021_mnras}. The average burst spectrum from the burst storm shows two $\sim100$\,MHz features (Figure\,\ref{fig:spectra}), reminiscent of the spectral structure in the \citet{majid_2021_apjl} \frb\ burst, and unlike typical repeater spectra. However, it has been shown that narrow-band FRB~20121102A bursts exhibit preferred frequencies, consistent on timescales of days \citep{gourdji_2019_apjl,hewitt_2022_mnras}. 

\subsection{Burst wait times}
\label{sec:discwait}
We find 3 peaks in the wait time distribution of bursts from \frb\ (Figure~\ref{fig:wait}). The main peak at $\sim 25$\,s simply reflects the overall burst rate, where on relatively short timescales of $< 1$\,hr we find the wait times to be reasonably well modelled by a Poissonian process. 

We also find a secondary peak in the burst wait times of \frb\ at $\sim 1$\,s. This is reminiscent of the secondary, shorter-timescale wait time peaks seen for FRB~20121102A \citep{gourdji_2019_apjl,li_2021_natur,hewitt_2022_mnras,jahns_2023_mnras} and FRB~20201124A \citep{xu_2022_natur}, though those sources both show such a peak at $\sim 30$\,ms, roughly a 50$\times$ shorter timescale. These secondary wait-time peaks demonstrate that once a burst has occurred, it is more likely to detect a second or third burst in short succession. This deviates from the general Poisson wait-time distribution. We suggest that these secondary wait-time peaks represent a timescale on which repeated burst emission can occur, and that this could be related to the overall physical size in which burst emission can be generated around the central engine as well as the timescale on which perturbations traverse this region. If so, then the much longer $\sim 1$\,s timescale of \frb\ could indicate a much larger overall emission region, or slower propagation of disturbances, compared to the $\sim 30$\,ms timescales that are observed for FRB~20121102A and FRB~20201124A. While the secondary wait-time peak of \frb\ is $\sim50\times$ longer in duration compared to FRB~20121102A and FRB~20201124A, its bursts are typically $\sim30\times$ shorter. It is worth considering whether these timescales are related.

Lastly, two of the bursts we detect from \frb\ show sub-millisecond separations between sub-bursts, and this suggests that there may also be a tertiary wait-time peak on this timescale. Other repeaters have also shown a characteristic spacing of wait-times between sub-bursts on timescales of roughly milliseconds (e.g., \citealt{hessels_2019_apjl}). This timescale may reflect the microphysics related to the coherent emission process: e.g., the interplay between charge bunching and radiative feedback \citep{lyutikov_2021_apj}. Some authors have also interpreted the quasi-periodic spacing of FRB sub-bursts in the context of outward propagating plasma oscillations \citep{sobacchi_2021_mnras}.

\subsection{Periodicity constraints}

We searched for a strict periodicity in the arrival times of the bursts, focusing on the relatively large sample of 53 bursts from the 2022 January 14 storm. Given the short burst durations of typically $\sim 100$\,$\upmu$s, our analysis should be sensitive to rotational periods of $\sim 1$\,ms or longer, if the bursts are clustered in rotational phase. Note that the methods we used are sensitive to bursts occurring in multiple rotational phase windows, as is sometimes seen from pulsars and radio-emitting magnetars \citep{camilo_2006_natur}. The sample of 53 bursts from the 2022 January 14 observation shows no statistically significant evidence for a short-duration period in the burst arrival times, and hence we conclude that --- if \frb\ is a rotating object with a period between 1\,ms and $25$\,s --- the bursts are roughly evenly distributed in rotational phase. This is consistent with the lack of detectable periodicity in other repeaters \citep{gourdji_2019_apjl,li_2021_natur,hewitt_2022_mnras}.

Nonetheless, we caution that a larger burst sample may reveal a more subtle clustering of bursts in rotational phase, and that clustering could potentially be time-variable: e.g., radio-emitting magnetars show evolving pulse profiles that can be stable on timescales of days to weeks or longer \citep{camilo_2006_natur}. The lack of observable periodicity distinguishes \frb\ from known giant pulse emitters and suggests that the emission region changes chaotically between the bursts.

We also searched the collection of known bursts from all our observations to see if there is evidence for periodicity in \frb's activity rate. Such searches are motivated by the well-established 16.3-day periodicity of FRB~20180916B \citep{chime_2020_natur_582} and the candidate 160-day periodicity of FRB~20121102A \citep{rajwade_2020_mnras,cruces_2021_mnras}. Though we find a hint for a 12.5-day period, a factor of a few more cycles are needed to ascertain if this will become statistically significant, or not.

\section{Conclusions \& future work}
\label{sec:conclusion}
We have presented the first-known `burst storm' from \frb, in which 53 bursts were detected in $< 1$\,hr of observation with the 100-m Effelsberg telescope on 2022 January 14. We characterise this event as a burst storm because of the high and rapidly varying burst rate, which is reminiscent of the X-ray burst storms seen from magnetars. The energy distribution of the burst storm is a steep power law ($\alpha = 2.39\pm0.12$ above a fluence threshold of 0.16\,Jy\,ms). We used these closely spaced bursts to search for a strict periodicity in the arrival times, but find no such signal, consistent with other known repeating FRBs. The burst wait times do, however, show a secondary peak at $\sim1$\,s. This is reminiscent of the secondary wait-time peaks seen for two other repeating FRBs on significantly shorter timescales of $\sim30$\,ms. The secondary wait time peak may represent a characteristic timescale related to the overall size of the system.

We note that a survey with a factor of 2 or 4 lower sensitivity than the Effelsberg observations presented here would have detected approximately 42 and 19 of the 53 bursts, respectively, and therefore also would have classified this event as a burst storm. Considering the instrumental sensitivity dependence on classifying burst storm events will be important in the future for determining the rate of such events from \frb\ or from other repeating FRBs. 

We also present an additional 7 bursts, which were detected in 4 other observing sessions in 2022 January through March. During these observations, the lower and more stable burst rate suggests that the source was in a different state compared to the 2022 January 14 storm. We used these and other observations, including those with non-detections, to search for periodic activity but find only tentative evidence for a 12.5-day period.

The observational record to date demonstrates that the DM of \frb\ is highly stable, unlike some other repeaters, and that its bursts are characteristically $\sim30\times$ shorter in duration and $\sim100\times$ less luminous compared to other known repeaters. Nonetheless, the narrow-band emission and time-frequency drift (`sad trombone' effect) seen from \frb\ provide important observational links to other repeaters. By comparing our $1.2-1.6$-GHz burst sample to that detected from $400-800$\,MHz by CHIME/FRB \citep{bhardwaj_2021_apjl}, we find that the average spectral index is quite steep: $\beta=2-3$ (note the caveat to this statement discussed in Section\,\ref{sec:subenerg}), indicating that there are good prospects for detecting \frb\ at very low frequencies ($<400$\,MHz).

We show that \frb\ bursts typically have fast rise ($\sim100$\,$\upmu$s) and slower decay ($\sim200$\,$\upmu$s), but that this asymmetry is not due to scattering. Whereas previous observations have demonstrated (sub-)microsecond structure in the bursts from \frb, we find that such short timescales occur rarely (or that the microstructure is typically bunched in time, as opposed to being resolvable). Additionally, searches of our voltage data using coherent dedispersion and a time resolution of $1.28$\,$\upmu$s led to the discovery of only one additional, low-S/N  burst compared to our initial searches at $40.96$\,$\upmu$s time resolution. This burst is the narrowest in the sample presented here, with a temporal scale of $\sim14$\,$\upmu$s. We find no bursts with durations $<10$\,$\upmu$s. 

Given the observations to date, we suggest that the most urgent avenues for future observational investigation are:

\begin{itemize}

    \item Detect the source both below 400\,MHz and above 3\,GHz to establish whether the burst widths are frequency-dependent.

    \item Continue the observational timeline in order to establish or rule out periodic variations in the source's activity.
    
    \item Use future detections to further constrain subtle DM and RM variations.

\end{itemize}

From the point-of-view of understanding the nature of \frb, we encourage theorists to focus on a self-consistent picture that explains the: i. low burst luminosities, ii. short burst durations, iii. $\sim1$-s secondary wait-time peak and iv. steep average burst spectrum. Each of these quantities deviates quite significantly from other known repeaters, and it may be possible to link these burst properties to those of the central engine (e.g., its rotation rate, magnetic field strength, age, etc.).

\section*{Acknowledgements}
K.~N. would like to thank P.~Chawla and K.~Rajwade for helpful discussions about the analysis presented in this work.

This paper is based on observations with the 100-m telescope of the MPIfR (Max-Planck-Institut f\''ur Radioastronomie) at Effelsberg.

Research by the AstroFlash group at University of Amsterdam, ASTRON and JIVE is supported in part by an NWO Vici grant (PI Hessels; VI.C.192.045).

This project has received funding from the European Union's Horizon 2020 research and innovation programme under grant agreements No 730562 (RadioNet) and 101004719 (OPTICON-RadioNet Pilot).


B.~M. acknowledges support from the Spanish Ministerio de Econom\'ia y Competitividad (MINECO) under grants AYA2016-76012-C3-1-P, FPA2017-82729-C6-2-R, and MDM-2014-0369 of ICCUB (Unidad de Excelencia ``Mar\'ia de Maeztu'')

\section*{Data Availability}
The filterbank data containing the FRBs presented in this work are available via Zenodo, at \href{https://doi.org/10.5281/zenodo.7555187}{https://doi.org/10.5281/zenodo.7555187}. Requests for other materials should be addressed to the lead author.



\bibliographystyle{mnras}



\bsp	
\label{lastpage}
\appendix
\section{Additional figures}
\setcounter{figure}{0}
\renewcommand{\thefigure}{A\arabic{figure}}
\begin{figure*}
\resizebox{\hsize}{!}
        {\includegraphics[trim=0cm 0cm 0cm 0cm, clip=true]{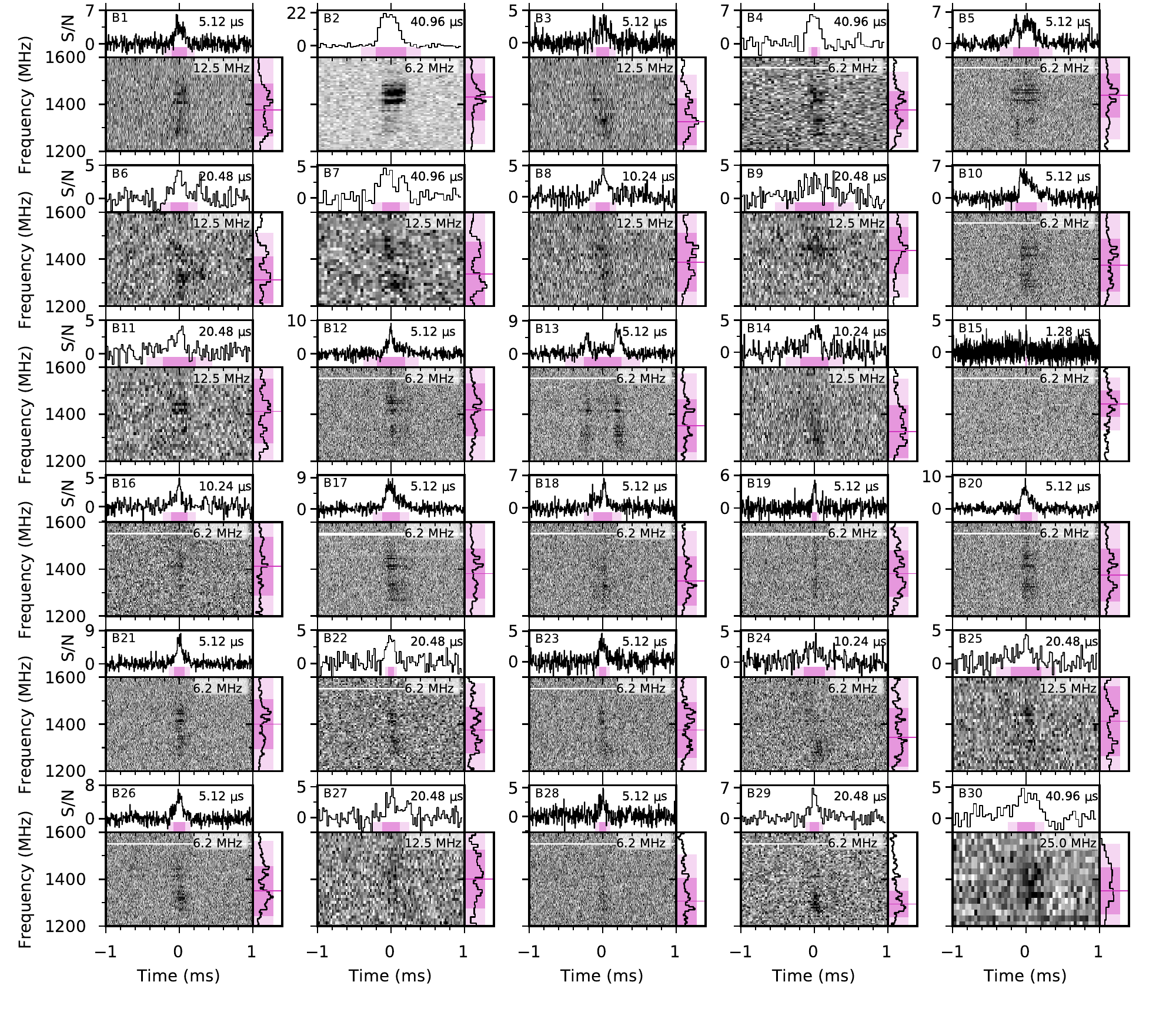}}
  \caption{(a) Frequency-averaged burst profiles, dynamic spectra and time-averaged spectra for all of the \frb\ bursts presented in this work, ordered according to their arrival time. The filterbank data plotted is created by channelising the raw voltages to a resolution of 5.12\,$\upmu$s and 0.1953\,MHz in time and frequency, respectively. The exception to this is B15, which was discovered in the microsecond search and the filterbank data is created with a resolution of 1.28\,$\upmu$s and 0.7813\,MHz (see Figure\,\ref{fig:somebursts} for a zoom-in panel of burst B15). Both the time and frequency have been downsampled by various factors for plotting, with the resolutions being shown on the burst profile and dynamic spectra, respectively. The data are coherently (within frequency channels) and incoherently (between channels) corrected for dispersion using a DM of 87.7527\,pc\,cm$^{-3}$ \citep{nimmo_2022_natas}. The coloured bars highlight the 1-$\sigma$ (dark) and 2-$\sigma$ (light) error regions on the burst extent in both frequency and time, where the different colours represent different observing epochs: 2022 January 14 (purple), 2022 January 17 (blue), 2022 February 21 (grey), 2022 February 23 (green) and 2022 March 17 (yellow). Time\,$=0$ represents the burst centroid in time, and the horizontal coloured line in the burst spectrum represents the frequency centroid. Frequency channels that have been masked due to RFI have been omitted from the dynamic spectra (horizontal white lines). The burst profile is created by averaging the dynamic spectrum within the 2-$\sigma$ burst extent in frequency, and similarly the burst spectrum is created by averaging the dynamic spectrum within the 2-$\sigma$ burst extent in time.}
     \label{fig:fullburst}
\end{figure*}
\setcounter{figure}{0}
\renewcommand{\thefigure}{A\arabic{figure}}

\begin{figure*}
\resizebox{\hsize}{!}
        {\includegraphics[trim=0cm 0cm 0cm 0cm, clip=true]{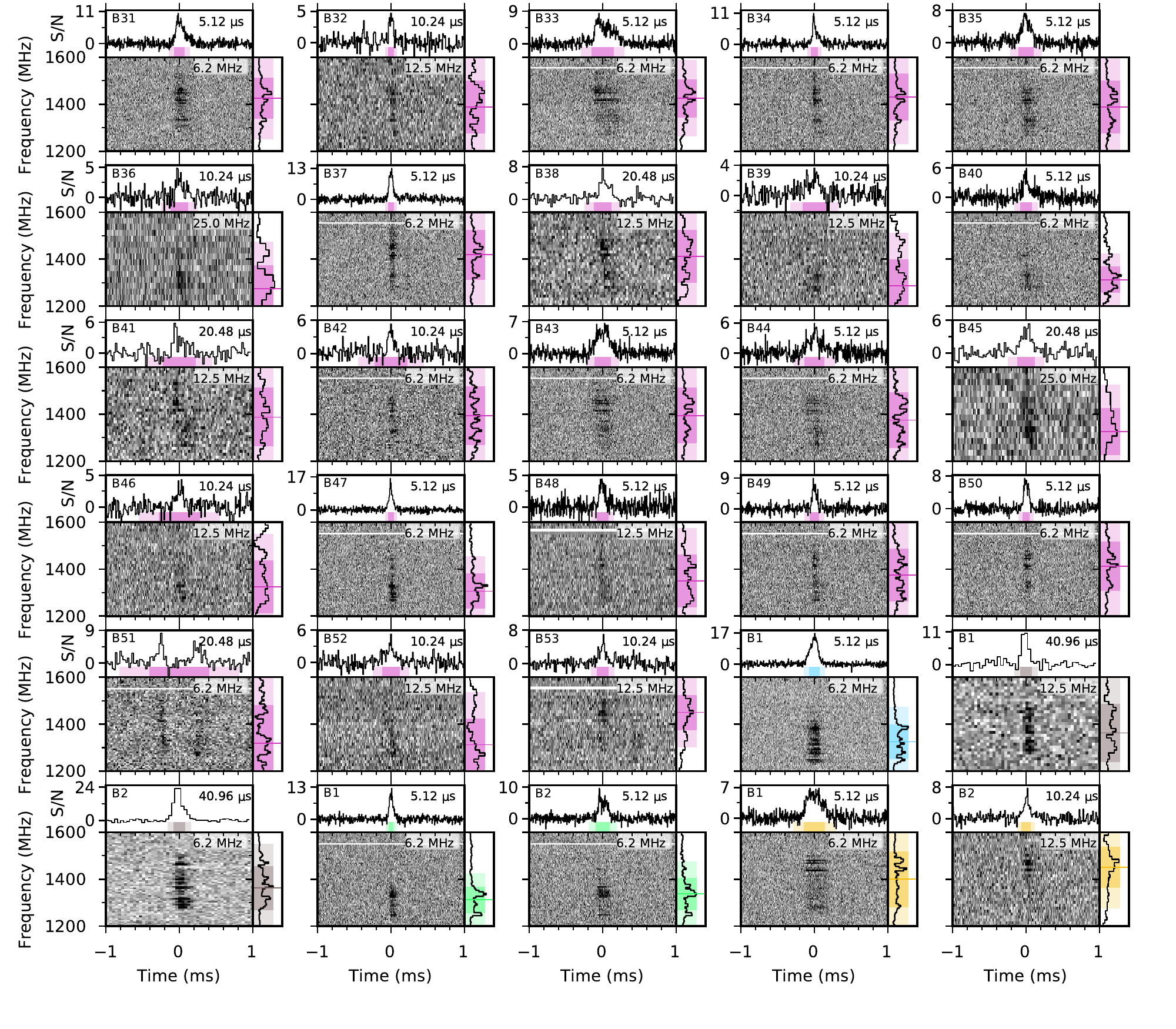}}
  \caption{(b) Continuation of Figure\,\ref{fig:fullburst}(a).}
\end{figure*}

\begin{figure*}
\centering
        {\includegraphics[height=200mm,trim=0.8cm 0cm 0cm 0cm, clip=true]{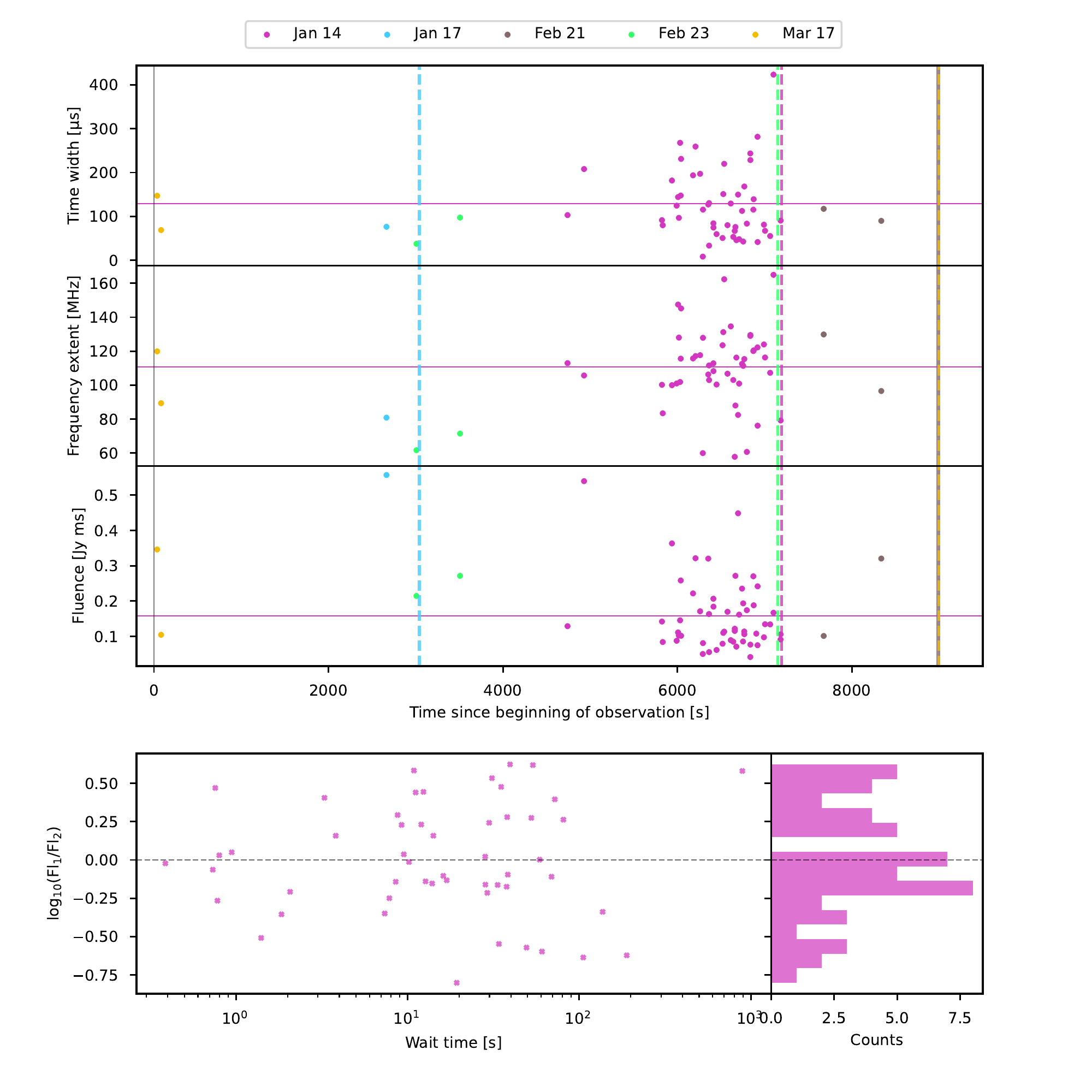}}
  \caption{Top sub-figure: Burst widths in time (top panel), bandwidths (middle panel) and fluences (bottom panel), as a function of time relative to the beginning of the observing epoch. As shown in the legend, the colours represent the different observing epochs. The vertical black line represents the start of the observations, and the coloured vertical dashed lines indicate the end of each observation. The horizontal purple line represents the mean value for the 2022 January 14 observation (burst storm). Bottom sub-figure: The ratio of burst fluences of consecutive bursts on 2022 January 14 as a function of their time separation, with a histogram shown on the right. The black dashed line represents the divide between the fluence of the first burst (Fl$_{1}$) being lower than the fluence of the second burst (Fl$_{2}$), and vice versa. }
     \label{fig:props_time}
\end{figure*}

\begin{figure*}
\centering
        {\includegraphics[trim=0cm 0cm 0cm 1.2cm, clip=true]{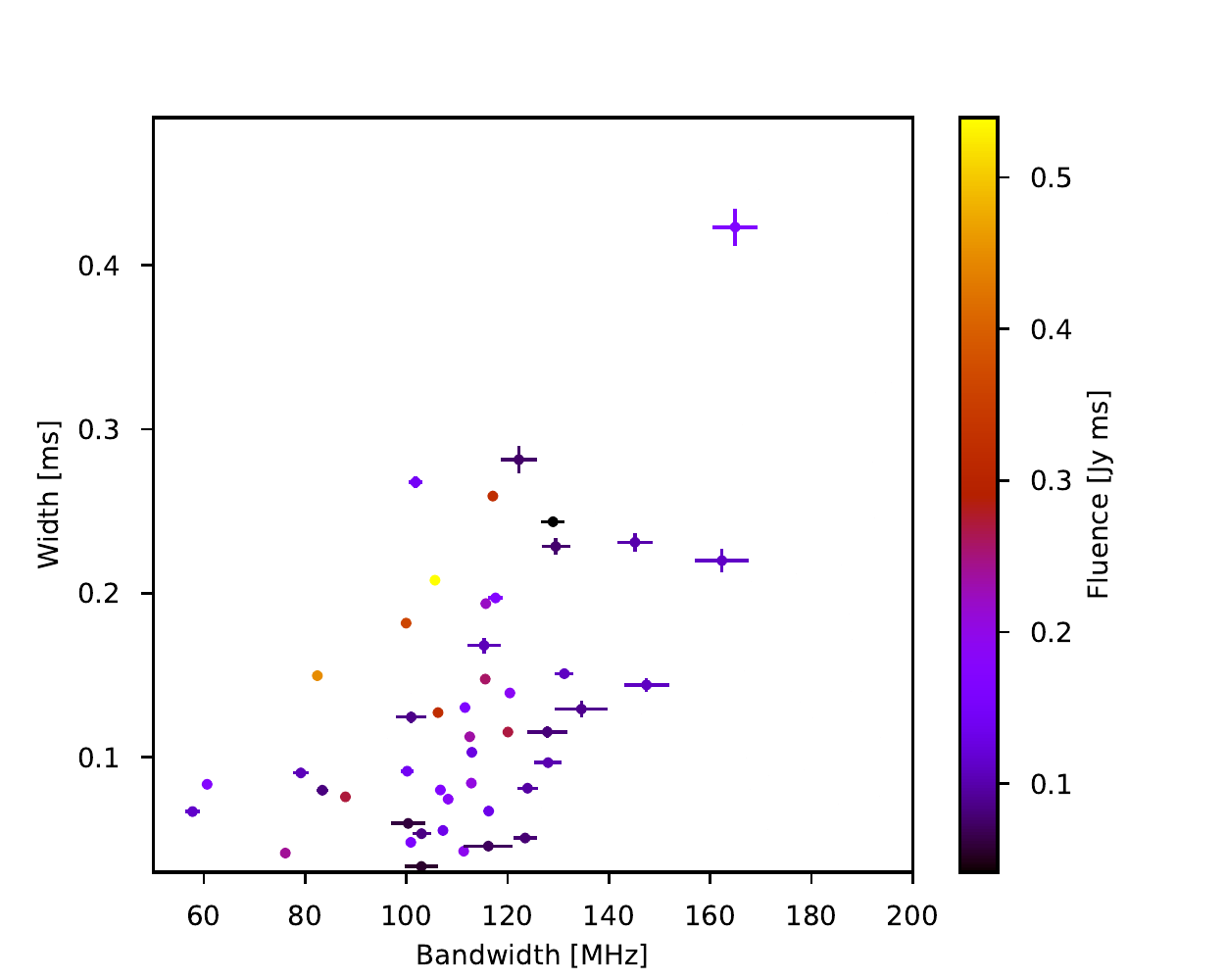}}
  \caption{Burst time duration versus frequency extent for bursts during the 2022 January 14 observation. The error bars are the $1\sigma$ fit errors from the 2-dimensional autocorrelation analysis described in the text, and reported in Table\,\ref{tab:burst_properties}. The colour scale represents the burst fluence. }
\end{figure*}

\begin{figure*}
\centering
        {\includegraphics[trim=0cm 0cm 0cm 0cm, clip=true]{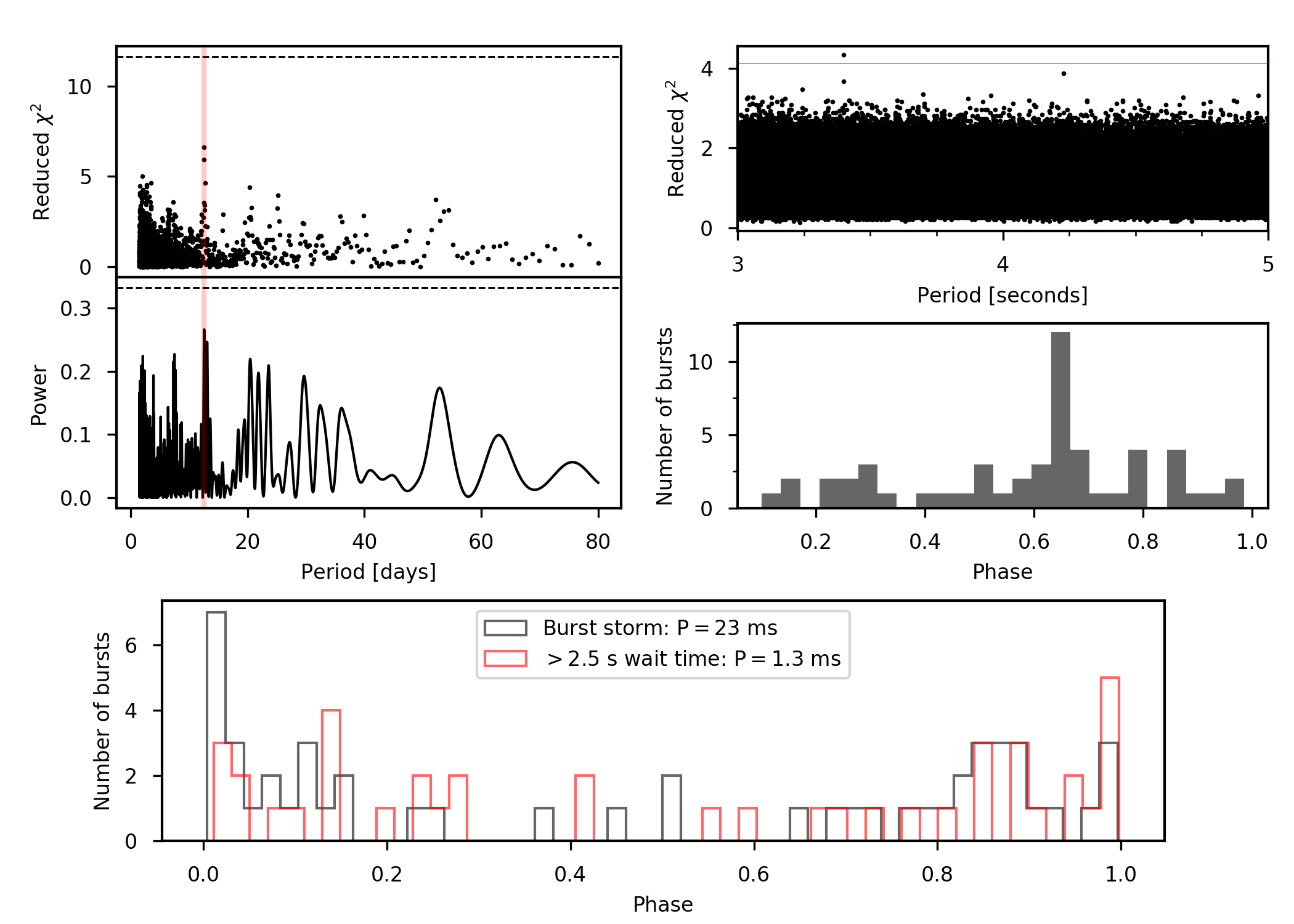}}
   \caption{\label{fig:period} Periodicity search results. Top-left sub-figure: The reduced $\chi^2$ for a uniform distribution of burst times as a function of trial activity period (top panel, using 3 bins) and the Lomb-Scargle periodogram (bottom panel) for the entire $>1$\,yr span of \frb\ observations. The $1-\sigma$ significance level is indicated by the dashed horizontal black line, calculated by bootstrapping 100 and 1000 trials for the reduced $\chi^2$ and Lomb-Scargle, respectively. The red vertical line highlights a candidate at 12.5\,days (although continuous monitoring of \frb\ is required to test the significance of this period). Top-right sub-figure: The reduced $\chi^2$ for a uniform distribution of burst times around the highest significance candidate for bursts detected during the 2022 January 14 observation (top panel, using 25 bins). The horizontal red line shows the 3$-\sigma$ significance, calculated from the $\chi^2$ survival function. The highest significance candidate (period 3.40\,ms) is barely $3\sigma$. Folding the burst ToAs using this period results in bursts arriving across all phases (bottom panel). Bottom sub-figure: The distribution of burst phases assuming the period measured from the brute-force search of an integer divisor of the burst ToAs. See Section\,\ref{sec:period} for details of the analysis resulting in this figure.  }
\end{figure*}

\end{document}